%% file: main.tex
\documentclass{lmcs}

\usepackage{hyperref}

\input{packages}

\input{macros}
\usepackage{algorithm}
\usepackage[noend]{algpseudocode}
\usepackage{comment}
\usepackage{amssymb}
\usepackage{tabularx}
\usepackage{pgffor}
\usepackage{tikz}
\usetikzlibrary{automata, positioning, arrows, graphs, quotes, calc, angles, shapes.multipart, trees, patterns, decorations.pathreplacing, decorations.markings}
\usepackage{placeins}
\usepackage{listings}

\newcommand{\para}[1]{\smallskip\noindent{\bf#1}}

\theoremstyle{plain}
\newtheorem{theorem}[thm]{Theorem}
\newtheorem{lemma}[thm]{Lemma}

\newtheorem{claim}[thm]{Claim}

\theoremstyle{definition}
\newtheorem{definition}[thm]{Definition}
\newtheorem{example}[thm]{Example}
\newtheorem{remark}[thm]{Remark}

\title{Sound Value Iteration for Simple Stochastic Games}

\thanks{This research was supported in part by the DFG project GOPro (427755713, Kretinsky), the MUNI Award in Science and Humanities (MUNI/I/1757/2021, Kretinsky), the EU Horizon Europe Grant (101171844, Kretinsky), the Marie Skłodowska-Curie Grant (101034413, Weininger), and the ERC Starting Grant DEUCE (101077178, Weininger).}

\author[M.~Azeem]{Muqsit Azeem\lmcsorcid{0000-0003-4532-8344}}[a]
\author[J.~K{\v{r}}et\'insk\'y]{Jan K{\v{r}}et\'insk\'y\lmcsorcid{0000-0002-8122-2881}}[b,a]
\author[M.~Weininger]{Maximilian Weininger\lmcsorcid{0000-0002-0163-2152}}[a,c]

\address{Technical University of Munich, Germany}
\email{muqsit.azeem@tum.de}

\address{Masaryk University, Brno, Czech Republic}
\email{jan.kretinsky@fi.muni.cz}

\address{Ruhr University Bochum, Germany}
\email{maximilian.weininger@ruhr-uni-bochum.de}

\begin{document}

	\begin{abstract}
		Algorithmic analysis of Markov decision processes (MDP) and stochastic games (SG) in practice relies on value-iteration (VI) algorithms. Since the basic version of VI does not provide guarantees on the precision of the result, variants of VI have been proposed that offer such guarantees.
		In particular, sound value iteration (SVI) not only provides precise lower and upper bounds on the result, but also converges faster in the presence of probabilistic cycles.
		Unfortunately, it is neither applicable to SG, nor to MDP with end components.
		In this paper, we extend SVI and cover both cases.
		The technical challenge consists mainly in proper treatment of end components, which require different handling than in the literature.
		Moreover, we provide several optimizations of SVI.
		Finally, we also evaluate our prototype implementation experimentally to confirm its advantages on systems with probabilistic cycles.
		
	\end{abstract}
	
	\keywords{Stochastic Games, Value Iteration, Verification}
	
	\maketitle
    
	\input{1_intro}

\input{2_prelims}

\input{3_svi-no-ec}
	\input{4_svi-with-ec}
	\input{5_topological}
	\input{6_experiments}

	\input{7_conclusion}	
	\input{8_appendix}

	\bibliographystyle{alphaurl}
	\bibliography{ref}

\end{document}

%% file: packages.tex
\usepackage{amsmath,amssymb,mathtools}
\usepackage{algorithm}
\usepackage{algorithmicx}
\usepackage[noend]{algpseudocode}
\usepackage{booktabs}
\usepackage{tikz}
\usepackage{nicefrac}
\usepackage{xparse}
\usepackage{xfrac}
\usepackage{multirow}
\usepackage{subcaption}
\usepackage{svg}
\usepackage{booktabs}
\PassOptionsToPackage{pdfstartview=FitH, pdfpagemode=UseNone}{hyperref}
\usepackage{hyperref}
\usepackage{xspace}
\usepackage{xcolor}
\usepackage{tikz}\usetikzlibrary{automata, positioning, arrows}
\tikzset {
	->,
	>=stealth',
	node distance=3cm,
	every state/.style={thick, fill=gray!10},
	initial text=$ $,
}

%% file: macros.tex
\usetikzlibrary{arrows,positioning,shapes,angles,arrows.meta,shapes.multipart,colorbrewer,backgrounds}
\usepackage{forest}

\definecolor{dartmouthgreen}{rgb}{0.05, 0.5, 0.06}
\definecolor{firebrick}{rgb}{0.7, 0.13, 0.13}

\definecolor{C-3-1}{RGB}{27,158,119}
\definecolor{C-3-2}{RGB}{217,95,2}
\definecolor{C-3-3}{RGB}{117,112,179}

\colorlet{disabled}{lightgray}
\colorlet{statecolor}{black}
\colorlet{rewardcolor}{C-3-2}
\colorlet{actioncolor}{black}
\colorlet{probcolor}{black}
\colorlet{highlightcolor}{C-3-1}

\tikzset{
	every picture/.style={
		>=stealth,
	},
	every edge/.append style={
		thick
	},
	highlight/.style={draw=rewardcolor,thick},
	state/.style={minimum width=0.8cm,minimum height=0.8cm,inner sep=1pt,rectangle,rounded corners,statecolor,draw},
	triangle/.style={regular polygon, regular polygon sides=3},
	max vertex/.style={state,path picture={\draw[gray,rounded corners=0,fill,fill opacity=0.2,thick] (path picture bounding box.south west) -- (path picture bounding box.north) -- (path picture bounding box.south east) -- cycle;}},
	min vertex/.style={state,path picture={\draw[gray,rounded corners=0,fill,fill opacity=0.2,thick] (path picture bounding box.north west) -- (path picture bounding box.south) -- (path picture bounding box.north east) -- cycle;}},
	goal/.style={state,font=\textsf{f},dartmouthgreen,fill=dartmouthgreen!15},
	sink/.style={state,font=\textsf{z},firebrick,fill=firebrick!15},
	action/.style={font=\small,inner sep=0pt,outer sep=3pt},
	reward/.style={inner sep=0pt,outer sep=3pt},
	actionnodelarge/.style={draw=actioncolor,rectangle,rounded corners=2pt},
	actionnode/.style={circle,draw=black,fill=black,minimum size=1mm,inner sep=0,outer sep=0,color=actioncolor},
	actionedge/.style={-,draw,color=actioncolor},
	prob/.style={font=\scriptsize,inner sep=0pt,outer sep=2pt,color=probcolor},
	probedge/.style={->,draw,color=probcolor},
	directedge/.style={->,draw,color=actioncolor}
}

\renewcommand{\circ}{\bigcirc}

\newcommand{\Naturals}{\mathbb{N}}
\newcommand{\Reals}{\mathbb{R}}
\newcommand{\Rationals}{\mathbb{Q}}
\newcommand{\abs}[1]{\lvert #1 \rvert}
\DeclareMathOperator*{\arginf}{arg\, inf}
\DeclareMathOperator*{\argmax}{arg\, max}

\newcommand{\eqdef}{\vcentcolon=}
\newcommand{\defeq}{=\vcentcolon}

\newcommand{\distribution}{\delta}
\newcommand{\Dist}{\mathsf{Dist}}

\newcommand{\post}{\mathsf{Post}}

\renewcommand{\Box}{\vartriangle}
\renewcommand{\circ}{\triangledown}
\newcommand{\G}{\mathcal{G}}

\newcommand{\states}{\mathsf{S}}
\newcommand{\minStates}{\states_{\circ}}
\newcommand{\maxStates}{\states_{\Box}}
\newcommand{\initstate}{s_{0}}

\newcommand{\act}{\mathsf{A}}
\newcommand{\Av}{\mathsf{Av}}
\newcommand{\trans}{\delta}

\newcommand{\sink}{\mathsf{z}}
\newcommand{\sinks}{\mathsf{Z}}
\newcommand{\target}{\mathsf{f}}
\newcommand{\targets}{\mathsf{F}}

\newcommand{\Bop}{\mathcal{B}}

\newcommand{\arbsigma}{\sigma^{\mathsf{arb}}}
\newcommand{\arbtau}{\tau^{\mathsf{arb}}}
\newcommand{\straas}{\stratsMax}

\newcommand{\stratsMax}{\Sigma_\Box}
\newcommand{\stratsMin}{\Sigma_\circ}
\newcommand{\Paths}{\mathsf{Paths}}

\newcommand{\reach}{\lozenge}
\newcommand{\stay}{\square}
\newcommand{\probability}{\mathcal{P}}
\newcommand{\val}{\mathsf{V}}
\newcommand{\lb}{\mathsf{L}}
\newcommand{\ub}{\mathsf{U}}

\newcommand{\unknown}{\states^?}
\newcommand{\svireach}{\newlink{def:reach-stay}{\mathsf{reach}}\xspace}
\newcommand{\svistay}{\newlink{def:reach-stay}{\mathsf{stay}}\xspace}

\newcommand{\svidecMax}{\mathsf{d_{u}}}

\newcommand{\glb}{\mathsf{l}}
\newcommand{\gub}{\mathsf{u}}
\newcommand{\mec}{\mathsf{Y}}

\newcommand{\chosen}{\mathsf{chosen}}
\newcommand{\decval}{\mathsf{decval}}
\newcommand{\deltastay}{\Delta_{\text{stay}}}
\newcommand{\deltareach}{\Delta_{\text{reach}}}

\newcommand{\bestExit}{\mathsf{bexit}}
\newcommand{\bestExitSet}{\mathfrak{B}}

\hypersetup{
	pdfstartview=FitH,  
	pdfpagemode=UseNone 
}

\definecolor{myblue}{RGB}{0, 102, 204}

\newcommand\newtarget[2]{ \phantomsection\hypertarget{#1}{#2}}

\newcommand\newlink[2]{{\protect\hyperlink{#1}{\color{myblue} #2}}}

\newcommand{\trap}{\newlink{def:trap}{\textsf{trap}}\xspace}

\newcommand{\opt}{\newlink{def:opt}{\textsf{opt}}\xspace}

\newcommand{\drawdummy}{\node[minimum size=0,inner sep=0]}

%% file: 1_intro.tex
\section{Introduction}

\para{Value iteration (VI)} \cite{bellman} is the practically most used method for reliable analysis of probabilistic systems, in particular Markov decision processes (MDP) \cite{Puterman} and stochastic games (SG) \cite{condonComplexity}.
It is used in the state-of-the-art model checkers such as Prism \cite{prism4} and Storm \cite{Storm} as the default method due to its better practical scalability, compared to strategy iteration or linear/quadratic programming \cite{practitionerguide,gandalf20}.
The price to pay are issues with precision.
Firstly, while other methods yield precise results (omitting
floating-point issues), VI converges to the exact result only in the limit.
Secondly, the precision of the intermediate iterations was until recently an open question.
Given the importance of reliable precision in verification, many recent works focused on modifying VI so that the imprecision can be bounded, yielding a stopping criterion.
Consequently, (i) the computed result is reliable, and (ii) the procedure can even terminate earlier whenever the desired precision is achieved.

Focusing on the reachability probabilities here, the methods for computing lower and upper bounds on this value include analysing \emph{end-components (ECs)} in graphs (\emph{bounded value iteration}~\cite{KKKW18} a.k.a \emph{interval iteration} \cite{hm18}, \emph{bounded real-time dynamic programming} \cite{atva}), finding augmenting-like paths (\emph{widest path} \cite{wp}), guessing the upper bound from the lower one (\emph{optimistic VI} \cite{OVI}), or \emph{sound VI} (SVI) \cite{DBLP:conf/cav/QuatmannK18}.

The main idea of SVI is to estimate the value by a geometric series describing the probabilities to (i)~reach the target in a given number of steps and (ii)~to remain undetermined in that number of steps.

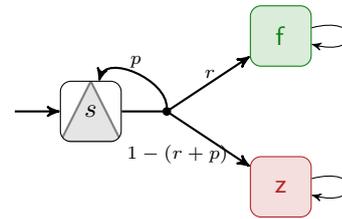
\begin{figure}
		\begin{tikzpicture}[scale=1.1, transform shape]
			\node[max vertex, ] at (2,0) (s) {$s$};
			\node[actionnode] at (3,0) (sc) {};
			\node[goal] at (4.5,1) (goal) {};
			\node[sink] at (4.5,-1) (sink) {};
			
			\draw[->, thick] (1,0) -- (s);
			\path[->]
			;
			\path[actionedge]
			(s) edge node[action,above] {} (sc)
			;
			\path[probedge]
			(sc) edge[out=90,in=75] node[prob,above] {$p$} (s)
			(sc) edge node[prob,above] {$r$} (goal)
			(sc) edge node[prob,right] {$~1-(r+p)$} (sink)
			;
			\draw[->]  (goal) to[loop right]  node [midway,anchor=west] {} (goal);
			\draw[->]  (sink) to[loop right]  node [midway,anchor=west] {} (sink);
		\end{tikzpicture}
	\caption{A Markov chain with an initial state $s$, a target state $\target$, and a sink/zero state $\sink$}
	\label{fig:3states}
	
\end{figure}

For instance, in the system depicted in Figure~\ref{fig:3states}, SVI essentially computes that within $n$ steps, the probability to reach the target is $\svireach_n=r+pr+p^2r+\cdots+p^{n-1}r$.
Moreover, since the probability to stay in undetermined states (neither target, nor sink) is $\svistay_n=p^n$, it deduces that the infinite-horizon reachability probability is in the interval $[\svireach_n, \svireach_n+\svistay_n]$.
Not only does the method provide such valid lower and upper bounds, but it also terminates (bounds are close enough to each other) comparatively faster when such probabilistic cycles are present, rendering  the usual VI very slow~\cite{DBLP:conf/cav/QuatmannK18}.

Unfortunately, SVI is only applicable to MDP without \emph{end components} (ECs).
While this is a common simplifying restriction (stochastic shortest path in MDP \cite{ssp}, stopping games \cite{shapley}), it is also very severe.
In this paper, \textbf{we extend SVI (i) to SG and (ii) to systems (both MDP and SG) with end-components (ECs)}.

First, we extend SVI to SG without ECs. 
While the algorithm is similar with analogously dual steps for the other player, the correctness is surprisingly not immediate. 
Indeed, the underlying optimization objective in SVI is not simple reachability but a non-trivial combination of step-bounded reachability of targets and of undetermined states.
This in turns makes these strategies require memory; the interaction of two memory-dependent strategies then convolutes constraints on the upper bound.

Second, we extend the approach to handle ECs.
A major portion of the literature on guaranteed-precision VI, i.e.~with lower and upper bounds, typically considers only the case without ECs \cite{DBLP:conf/cav/QuatmannK18}; for MDP, the ECs can be \emph{collapsed} before the analysis, avoiding the issue, but \emph{not so for SG}.
The main approach to treat ECs is \emph{deflating} \cite{KKKW18}, which adjusts the upper bound of each state to reflect the true probability of reaching the target \emph{outside} of the EC.
Since SVI does not work with the explicit upper bound, but rather with an implicit version given by other values (such as $\svireach_n$ and $\svistay_n$), deflating cannot be performed directly.
Consequently, (i) instead of a single (S)VI update, we shall have four different cases to handle; (ii)  the fundamental notion of the \emph{simple end component} (EC states with the same value), instrumental for other VI extensions, is too coarse; instead, we have to consider their yet smaller parts;
(iii) the convergence of the bounds does not follow directly from the literature.
In particular, instead of the easy, standard convergence of the lower bound as seen in all other VI variants, we need properties of the upper bound to argue about the convergence of the lower bound.
This constitutes the technically most involved part of the paper.

\para{Our contribution} can be summarized as follows: 
\begin{itemize}
	\item After providing an instructive description of SVI (in particular adding some aspects lacking in the original \cite{DBLP:conf/cav/QuatmannK18}),
	we generalize SVI to (i) stochastic games (SGs) and, crucially, (ii) systems with end components (ECs), thereby overcoming a fundamental limitation of SVI in the presence of ECs.
	\item Based on the performed in-depth analysis, we propose topological improvements of SVI (Section~\ref{sec:toplogical}) that replace global bounds with bounds over the relevant reachable part of the state space, avoiding global bottlenecks and enabling more informed updates.
	\item 
	Our prototypical implementation and experimental analysis illustrate that our extension preserves the main advantage of SVI (see Section~\ref{svi-good-example} for a concrete example)---fewer iterations are needed for systems with probabilistic cycles.
\end{itemize}

We emphasize that our contribution does not lie in providing a tool more efficient than other ones in the
literature, but rather in the theoretically involved proof and way how SVI can be extended to handle ECs.
Specifically, our contribution lies in highlighting the inductive, attractor-like structure of ECs (Section~\ref{ssec:bes}),
which provides a theoretical foundation for addressing them---the primary challenge with VI. Instead
of aiming to replace BVI---which handles ECs (i.e., sure cycles)---this work aims to complement it by
offering critical insights into the inherent drawbacks of VI, such as its slow convergence in the presence
of probabilistic cycles. Our insights pave the way for potential future enhancements to BVI through
the efficient probabilistic cycle handling introduced by SVI, which is currently missing in BVI. This
foundational step, though part of a longer-term effort, is pivotal for enabling subsequent advancements.

Compared to the conference version~\cite{gandalf25}, this paper provides a substantially extended and self-contained presentation of our results. In particular, we include full proofs of all main results, which were previously deferred to the appendix, and integrate them into the main exposition, building upon with now fully self-contained preliminaries. In Section~\ref{sec:decision-value}, we include an extensive derivation and explanation of the notion of \emph{decision value}~\cite{DBLP:conf/cav/QuatmannK18}, detailing on and streamlining the original exposition of \cite{DBLP:conf/cav/QuatmannK18}, additionally amending it with a detailed example. We also refine the presentation of the algorithm for handling end components, including a detailed treatment of end-component decomposition and delay actions, together with a discussion of correctness. The experimental evaluation is now integrated, illustrating the behavior of the approach.

The remainder of the paper is structured as follows. Section~\ref{sec:prelims} introduces the necessary preliminaries, including SVI for Markov chains. Section~\ref{sec:svi-ec-free} extends this to stochastic games without ECs. Section~\ref{sec:algorithm-with-ec} lifts this restriction by handling ECs. Section~\ref{sec:toplogical} presents topological improvements of SVI. Section~\ref{sec:experiments} reports on the experimental evaluation. Finally, Section~\ref{sec:conc} concludes.

%% file: 2_prelims.tex
\section{Preliminaries}
\label{sec:prelims}

We briefly recall stochastic games and introduce our notation.
For an extensive introduction, we refer the interested reader to~\cite[Chapter 2]{MaxiThesis}.

\begin{definition}[Stochastic Game (SG)~\cite{DBLP:conf/dimacs/Condon90}]
	A (turn-based simple) stochastic game is $\G = \langle \states, \maxStates, \minStates, \act, \Av, \delta, \targets \rangle$ where $\states = \maxStates \uplus \minStates$ is a finite set of \emph{states} partitioned into \emph{Maximizer} ($\maxStates$) and \emph{Minimizer} ($\minStates$) states, 
	$\act$ is a finite set of \emph{actions}, the function $\Av: \states \to 2^\act \setminus \{\emptyset\} $ assigns \emph{available} actions to each state, $\delta:\states \times \act \to \Dist(\states)$ is the \emph{transition function} mapping state-action pairs to probability distributions over $\states$, and $\targets\subseteq \states$ is the \emph{target} set.
\end{definition}

An SG with $\minStates=\emptyset$ or $\maxStates=\emptyset$ is called a \emph{Markov decision process (MDP)}. 
For $s \in \states,a \in \Av(s)$, the set of \emph{successors} is $\post(s,a) := \{s' \ | \ \distribution(s,a,s') >0 \} $.
For a set of states $T \subseteq \states$, we use $T_\Box = T \cap \maxStates$ to denote all Maximizer states in $T$, and similarly $T_\circ$ for Minimizer.

The semantics is defined as usual by means of paths and strategies.
An infinite path is a sequence $s_0a_0s_1a_1\ldots \in (\states \times \act)^\omega$ that is consistent with the SG, i.e.\ for each $i \in \mathbb{N}_0$ we have $a_i \in \Av(s_i)$ and $s_{i+1} \in \post(s_i, a_i)$. A history (also called finite path) is a finite prefix of such an infinite path, i.e.\ an element of $(\states\times\act)^* \times \states$ that is consistent with the SG. 
For a history $\rho = s_0a_0s_1a_1 \ldots s_k$, let $\rho_i\vcentcolon=s_i$ denote the $i$-th state in a path and $\abs{\rho} = k$ denote the length of the history (only counting actions; intuitively: the number of steps taken).

A strategy is a (partial) function $\pi \colon (\states\times\act)^* \times \states ~\to~ \act$ mapping histories to actions. 
A Maximizer strategy is only defined for histories that end in a Maximizer state $s\in\maxStates$, and dually for Minimizer.
We denote by $\stratsMax$ and $\stratsMin$ the sets of all Maximizer and Minimizer strategies and use the convention that $\sigma\in\stratsMax$ and $\tau\in\stratsMin$.
Complementing an SG $\G$ with a pair of strategies (essentially resolving all non-determinism) induces a Markov chain $\G^{\sigma,\tau}$ with the standard probability space $\probability_{s,\G}^{\sigma,\tau}$  in state $s$, see~\cite[Sec. 2.2]{MaxiThesis}.

The \emph{reachability objective $\reach\targets$} is the set of all infinite paths containing a state in $\targets$. Formally, let $\Paths$ be the set of all infinite paths. Then \[\reach\targets = \{\rho\in\Paths \mid \exists i\in\Naturals. \rho_i \in \targets\}.\] 
Later, we also use the step-bounded variant 
\[\reach^{\leq k}\targets = \{\rho\in\Paths \mid \exists i\in\Naturals. \rho_i \in \targets \wedge i\leq k\},\] 
and the dual step-bounded staying objective 
\[\stay^{\leq k}\unknown = \Paths \setminus \Big( \reach^{\leq k}(\states\setminus\unknown) \Big).\]
The \emph{value} of a state $s$ is the probability that from $s$ the target is reached under an optimal play by both players: $	\val(s) = \sup_{\sigma\in\stratsMax}\inf_{\tau\in\stratsMin} \probability_s^{\sigma,\tau} (\reach \targets) = \inf_{\tau\in\stratsMin}\sup_{\sigma\in\stratsMax} \probability_s^{\sigma,\tau} (\reach \targets)$; the order of choosing the strategies does not matter~\cite{condonComplexity}.
The function $\val: \states \to \mathbb{R}$ maps every state to its value. 
Functions $f_1, f_2: \states \to \mathbb{R}$ are compared point-wise, i.e.\ $f_1 \leq f_2$ if for all $s \in \states$, $f_1(s) \leq f_2(s)$.
In this paper, we denote $f(s, a) = \sum_{s' \in \states} \distribution(s,a,s') \cdot f(s')$ for any $f : \states \rightarrow \mathbb{R}$, using notation overloading.

\subsection{Value iteration and bounded value iteration}
\label{sec:prelims:vi}

Before we define the algorithm of value iteration, we recall a useful partitioning of the state space~\cite{DBLP:conf/cav/QuatmannK18}.
We know $\targets$ is the set of target states; w.l.o.g., every target state is absorbing.
Similarly, we can identify by simple graph search the set $\sinks$ of \emph{sink states} from which there is no path to $\targets$.
The values of states in these sets are trivially 1 or 0, respectively.
Our algorithm focuses on computing the values of the remaining states in $\unknown = \states \setminus (\targets \cup \sinks)$.

The value function $\val : \states \to [0,1]$ is the least fixpoint of the following Bellman equations:
\begin{equation}
	\label{eqn:value-equation}
	\val(s) =
	\begin{cases}
		1 & \text{if } s \in \targets,\\
		0 & \text{if } s \in \sinks,\\
		\max_{a \in \Av(s)} \sum_{s' \in \states} \distribution(s,a,s') \cdot \val(s') & \text{if } s \in \maxStates,\\
		\min_{a \in \Av(s)} \sum_{s' \in \states} \distribution(s,a,s') \cdot \val(s') & \text{if } s \in \minStates.
	\end{cases}
\end{equation}

We define the Bellman operato $\Bop : (\states \to \mathbb{R}) \to (\states \to \mathbb{R})$ as:
\begin{equation}
	\label{eqn:bellman-operator}
	\Bop(f)(s) =
	\begin{cases}
		\max_{a \in \Av(s)} \sum_{s' \in \states} \distribution(s,a,s') \cdot f(s') & \text{if } s \in \maxStates,\\
		\min_{a \in \Av(s)} \sum_{s' \in \states} \distribution(s,a,s') \cdot f(s') & \text{if } s \in \minStates.
	\end{cases}
\end{equation}

Classical value iteration (VI, see~\cite{visurvey}) computes a sequence $\lb_i:\states\to\mathbb{R}$ of under-approximations.
It starts with
\[
\lb_0(s) =
\begin{cases}
	1 & \text{if } s \in \targets,\\
	0 & \text{otherwise}
\end{cases}
\]
and iteratively applies the Bellman operator, i.e., $\lb_{i+1} = \Bop(\lb_i)$.
These under-approximations converge (in the limit) to the value $\val$, which is the least fixpoint of $\Bop$.

While this approach is often fast in practice, it has the drawback that the imprecision $\val-\lb_i$ is unknown.
To address this, one can employ \emph{bounded value iteration} (BVI, also known as interval iteration~\cite{atva,hm18,KKKW18}).
It additionally computes an over-approximating sequence $\ub_i$ which allows to estimate the imprecision as $\ub_i(s)-\lb_i(s)$. 
Naively, the sequence is initialized as $\ub_0(s) = 0$ for all $s \in \sinks$ and $\ub_0(s) = 1$ for all other states, and then updated using Equation~\eqref{eqn:bellman-operator} again.
However, then $\ub_i$ converges to the \emph{greatest} fixpoint of Equation~\eqref{eqn:bellman-operator} which can differ from the least fixpoint $\val$; this happens if and only if \emph{end components} are present.
\begin{definition}[End component (EC)~\cite{dA97a}]
	\label{def:EC}
	A set of states $T$ with $\emptyset \neq T \subseteq \states$ is an \emph{end component} if there exists a set of actions $\emptyset \neq B \subseteq \bigcup_{s \in T}Av(s)$ such that:\\
	1. for each $s \in T$, $a \in B \cap \Av(s)$ we have $\post(s,a) \subseteq T$, \\
	2. for each $s, s' \in T$ there exists a path from $s$ to $s'$ using only actions in $B$.
	
	An end component $T$ is a \emph{maximal end component} (MEC) if there is no other EC $T'$ such that $T \subset T'$. Furthermore, a MEC $T$ is \emph{bottom} if for all $s \in T, a\in\act, s' \in \states \setminus T$ we have $\distribution(s,a,s')=0$.
\end{definition}

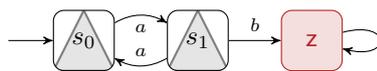
\begin{figure}[H]
	\begin{center}
		\begin{tikzpicture}[scale=1.2, transform shape]
		\drawdummy (init) at (0,0) {};
		\node[max vertex, ] (q) at (1,0) {$s_0$};
		\node[max vertex] (r) at (2.5,0) {$s_1$};
		\node[sink] (0) at (4,0)  {};
		\draw (init) to (q);
		\draw (q) to [bend left] node[below]{\scriptsize $a$} (r);
		\draw (r) to [bend left] node[above]{\scriptsize $a$} (q);
		\draw (r) to node[above]{\scriptsize $b$} (0);
		\draw (0) to[loop right]  node [midway,anchor=west] {} (0);
		\end{tikzpicture}
	\end{center}
	\caption{An SG with an EC $\{s_0,s_1\}$ and a sink $\sink$}
	\label{fig:2states}
\end{figure}

As an example, consider the MDP in Figure~\ref{fig:2states}. Here, any $\ub$ with $\ub(s_0)=\ub(s_1)$ and $\ub(\sink)=0$ is a fixpoint of Equation~\eqref{eqn:bellman-operator}.
More generally, on an EC in a maximizing MDP, one can create a fixpoint by assigning any number larger than the actual value to all states of the EC (and dually, a smaller number in a minimizing MDP).
In MDP, there is a rather simple solution: the entire EC can be  collapsed into a single state, keeping all the actions, but then removing all self-loops. This ensures a single fixpoint and, since all states in an EC have the same value as the collapsed state, it allows to obtain the values for the original MDP~\cite{atva,hm18}.
For SG, this solution is not applicable because states in an EC can have different values, see~\cite[Section 3]{KKKW18}.
The solution of~\cite{KKKW18} is to repeatedly identify subsets of ECs that possibly, given the current estimates, have the property that all states have the same value; then, the over-approximation in these subsets is conservatively lowered (\emph{``deflated''}) to the over-approximation of the best action \emph{exiting} the subset.
Relevant for our work is the fact that we need to decrease the over-approximation inside ECs to the estimate of the \emph{best exit}. Thus, we repeat here a (slightly adjusted) definition of best exit.

\begin{definition}[Best exit~\cite{KKKW18}]
	\label{def:bestExit}
	Given a set of states $T \subseteq \states$ and a function $f : \states \rightarrow \Rationals$, the best exit
	according to $f$ from $T$ is defined as:
	\begin{align*}
		\bestExit^{\Box}_f(T) &= 
		\displaystyle \argmax_{\substack{(s,a) \in T_\Box\times \Av(s)\\ \post(s,a) \nsubseteq T}} \ \sum_{s' \in \states} \distribution(s,a,s') \cdot f(s') 
	\end{align*}
\end{definition}

\subsection{Sound value iteration in Markov chains}

\label{sec:prelims:svi}

Sound value iteration (SVI,~\cite{DBLP:conf/cav/QuatmannK18}) was introduced as an alternative to BVI that is faster while still providing guarantees on its precision. 
It computes for increasing $k$ the probability of reaching the target set $\targets$ within $k$ steps, denoted $\probability(\reach^{\leq k} \targets)$, and the probability to stay within the unknown set $\unknown$ during the first $k$ steps, denoted $\probability(\stay^{\leq k} \unknown)$.
From these two sequences, it computes under- and over-approximations of the value utilizing three core ideas.
\begin{itemize}
	\item The value can be split into the probability to reach the target set \emph{within $k$ steps} plus the probability to reach the target set \emph{in strictly more than $k$ steps}~\cite[Lemma 1]{DBLP:conf/cav/QuatmannK18}.
	The former $\probability(\reach^{\leq k} \targets)$ is directly computed by the algorithm; the latter is approximated using the next two ideas.
	
	\item Reaching the target set in strictly more than $k$ steps can be decomposed as follows: first a path has to stay in $\unknown$ for $k$ steps and afterwards reach the target. As we do not know which state in $\unknown$ is reached after the $k$ steps, we estimate its value using a lower bound $\glb_k$ or an upper bound $\gub_k$ on the value over \emph{all states in $\unknown$}.
	Formally, given such a bound $\glb_k$, the probability to reach in strictly more than $k$ steps can be under-approximated as $\probability(\stay^{\leq k} \unknown)\cdot\glb_k$; dually, it can be over-approximated as $\probability(\stay^{\leq k} \unknown)\cdot\gub_k$~\cite[Proposition 1]{DBLP:conf/cav/QuatmannK18}.
	
	\item Such bounds $\glb_k$ and $\gub_k$ can be obtained from the $k$-step bounded probabilities $\probability(\reach^{\leq k} \targets)$ and  $\probability(\stay^{\leq k} \unknown)$~\cite[Proposition 2]{DBLP:conf/cav/QuatmannK18}.
	Intuitively, the highest possible value would be obtained if the state $\hat{s}\in\unknown$ with the highest value always reached itself after $k$ steps --- getting another try to obtain the highest possible value. The resulting value can be computed as $\sum_{i=0}^\infty \probability_{\hat{s}}(\stay^{\leq k} \unknown)^i\cdot\probability_{\hat{s}}(\reach^{\leq k} \targets)  $. 
	Writing this geometric series in closed form (assuming $\probability_{\hat{s}}(\stay^{\leq k} \unknown) < 1$), we obtain
	\[
	\sum_{i=0}^\infty \probability_{\hat{s}}(\stay^{\leq k} \unknown)^i\cdot\probability_{\hat{s}}(\reach^{\leq k} \targets)=
		\frac{\probability_{\hat{s}}(\reach^{\leq k} \targets)}{1-\probability_{\hat{s}}(\stay^{\leq k} \unknown)} \leq \max_{s\in\unknown} \frac{\probability_{s}(\reach^{\leq k} \targets)}{1-\probability_{s}(\stay^{\leq k} \unknown)} =\colon \gub_k
	\]	
	Dually, we can compute a lower bound $\glb_k$ by picking the state in $\unknown$ that minimizes the closed form of the geometric series.
	The following example illustrates the power of this way of choosing the lower and upper bound. Consider again the Markov chain of Fig.~\ref{fig:3states} and set $p=0.98$ and $r=0.01$. Then, while BVI requires 682 iterations to achieve a precision of $10^{-6}$, SVI takes precisely 1 iteration.
	SVI's behaviour is in fact independent of $p$ and $r$, whereas increasing $p$ also increases the number of iterations that BVI requires.
\end{itemize}

We restate~\cite[Theorem 1]{DBLP:conf/cav/QuatmannK18}, which summarizes the above insights and proves the correctness of the approach.

\begin{theorem}
	\label{thm:svi-mc} 
	Let $\mathcal{M}$ be a Markov chain with probability measure $\probability$ whose state space is partitioned into $\targets$, $\unknown$ and $\sinks$ as described above. 
	Let $k\geq0$ such that $\probability (\square^{\leq k} \unknown) < 1$ for all $s \in \unknown$. 
	Then, for every state $s\in\unknown$, its value $\val(s)=\probability_s(\reach \targets)$ satisfies the following:
	
	\[
		\probability_s(\reach^{\leq k} \targets) + \probability_s
		(\square^{\leq k} \unknown) \cdot \glb_k \leq \val(s) \leq \probability_s
		(\reach^{\leq k} \targets) + \probability_s (\square^{\leq k} \unknown)
		\cdot \gub_k
	\]
	
	where $\glb_k=\min_{s' \in \unknown} \frac{\probability_{s'}(\reach^{\leq k} \targets)}{1-\probability_{s'} (\square^{\leq k} \unknown)}$ and dually $\gub_k=\max_{s' \in \unknown} \frac{\probability_{s'}(\reach^{\leq k} \targets)}{1-\probability_{s'} (\square^{\leq k} \unknown)}$.
	
\end{theorem}

The termination of the algorithm relies on the assumption that the system is contracting towards $\targets\cup\sinks$; intuitively, this means that there are no sure cycles in $\unknown$ and thus a path has to leave $\unknown$ almost surely.
However, in MDP and SG there can be sure cycles, namely the ECs, which also hinder the convergence of BVI.  
In Section~\ref{sec:algorithm-with-ec}, we provide a way to deal with the presence of ECs when using SVI; note that neither the previous solution of collapsing~\cite{atva,hm18} nor of deflating~\cite{KKKW18} suffice.

However, before we resolve the problems caused by ECs, we first turn to a different complication in Section~\ref{sec:svi-ec-free}.
Our reasoning so far was limited to Markov chains with no non-determinism. 
It is non-trivial to extend it to MDP (and even more so to SG here) because there the computations additionally depend on the choice of strategy (strategies). In particular, to ensure correctness, the probability to reach within $k$ steps depends on the chosen strategies and need not be the optimal $k$-step value; moreover, the bounds cannot be computed using only the closed form of the geometric series.

%% file: 3_svi-no-ec.tex
\section{Sound value iteration for stochastic games without end components}\label{sec:svi-ec-free}

In this section, we provide the extension of SVI from MDP to SG without ECs. We extend it for the second player, building upon previous work~\cite{DBLP:conf/cav/QuatmannK18} by performing necessary dual operations.  
This section serves as an introduction to the basic algorithm, revisiting SVI concepts with additional examples, improved notation, and the part of the correctness proof that was missing in \cite{DBLP:conf/cav/QuatmannK18}. Additionally, this section prepares the reader for the more technical discussion on algorithms involving ECs in Section~\ref{sec:algorithm-with-ec}.

In order to lift Theorem~\ref{thm:svi-mc} from Markov chains to SG, we have to answer the following two questions:
\begin{enumerate}
	\item[1.] How do we pick the strategies $\sigma_k$ and $\tau_k$ that Maximizer and Minimizer use in the $k$-th iteration? (Section~\ref{sec:svi-strats})
	\item[2.] How do we correctly pick the lower bound $\glb_k$ and the upper bound $\gub_k$ on the value of all states in $\unknown$? We address this in Section~\ref{sec:no-ec-bounds}, building on the notion of decision values (Section~\ref{sec:decision-value}).
\end{enumerate}

Finally, Section~\ref{sec:algorithm-no-ec} combines the insights into the full algorithm.

\subsection{Computing the strategies}
\label{sec:svi-strats}
Following the idea of~\cite[Section 4.1]{DBLP:conf/cav/QuatmannK18}, we observe that for an optimal Maximizer strategy $\sigma^{\text{max}}$ and for all Minimizer strategies $\arbtau$ and a bound $\gub_k \geq \max_{s\in\unknown} \probability_s^{\sigma^{\max},\arbtau}(\reach \targets)$, we get the following chain of inequations:

\begin{align}
	\val = \sup_{\sigma\in\stratsMax}\inf_{\tau\in\stratsMin} \probability^{\sigma,\tau}(\reach \targets) &=
	\inf_{\tau\in\stratsMin} \probability^{\sigma^{\text{max}},\tau}(\reach \targets) \nonumber \\ &\leq \probability^{\sigma^{\text{max}},\arbtau}(\reach \targets) 
	\nonumber \\
	&\leq 
	\probability^{\sigma^{\text{max}},\arbtau}(\reach^{\leq k} \targets) + \probability^{\sigma^{\text{max}},\arbtau}(\stay^{\leq k} \unknown)\cdot \gub_k
	\nonumber \\
	&\leq \sup_\sigma \left(\probability^{\sigma,\arbtau}(\reach^{\leq k} \targets) + \probability^{\sigma,
		\arbtau}(\stay^{\leq k} \unknown)\cdot \gub_k\right)
	\label{eq:maxCorr}\hspace*{-4mm}
\end{align}
We can use the dual argument for a lower bound $\glb_k$, an arbitrary Maximizer strategy $\arbsigma$ and an optimal Minimizer strategy to obtain:
\begin{align}
	\label{eq:minCorr}
	\val \geq \inf_\tau \left(\probability^{\arbsigma,\tau}(\reach^{\leq k} \targets) + \probability^{\arbsigma,\tau}(\stay^{\leq k} \unknown)\cdot \glb_k\right)
\end{align}

Based on these inequalities, we can inductively construct strategies $\sigma_k,\tau_k$.%
\newtarget{def:opt}
In order to highlight the symmetry of the players, our formal definition uses a strategy $\pi$ of either Maximizer or Minimizer.
There are two differences between the players: the operator $\opt$ is $\max$ if $\pi\in\stratsMax$ and dually $\min$ if $\pi\in\stratsMin$ (with $\overline{\opt}=\min$ if $\opt=\max$ and vice versa) and the bound $\mathsf{b}_k$ is either $\gub_k$ or $\glb_k$, respectively.

Importantly, the strategies should not optimize ($k$-step) reachability, but the $k$-step objective $\probability^{\sigma_k,\tau_k}(\reach^{\leq k} \targets) + \probability^{\sigma_k,\tau_k}(\stay^{\leq k} \unknown)\cdot \mathsf b_k$, as suggested by Equations~\eqref{eq:maxCorr} and~\eqref{eq:minCorr}.

We construct the strategies recursively, building on the correctness of the strategies with shorter horizon. To achieve this, the $(k+1)$st strategy makes one optimal choice in the first step and then mimics the $k$th strategy, i.e.\ for a path $sa\rho$ starting with state $s$ and chosen action $a$, afterwards we have $\pi_{k+1}((sa)\rho)=\pi_k(\rho)$.
As base case, when $k=0$ we choose an arbitrary action, because all actions are 0-step optimal.
Formally,

\begin{equation}
		\label{eq:strats}
	\boxed{
		\begin{aligned}
					&\textsc{``$k$-step'' Strategies:}\\
			&\pi_0(\rho) \in \Av(\rho_0) \\	
			&\pi_{k+1}(\rho) \gets
			\begin{cases}
				\displaystyle
				\in \arg \opt_{a\in\Av(s)} \sum_{s' \in S} \distribution(s, a, s') \cdot \Big(\probability^{\sigma_k, \tau_k}_{s'}  (\reach^{\leq k}\targets) + \probability^{\sigma_k, \tau_k}_{s'}  (\stay^{\leq k}\unknown) \cdot \mathsf b_k \Big) &\mbox{if } \rho= s\in\states \\
				\pi_k(\rho') &\mbox{if } {\rho}=sa\rho'
			\end{cases}
		\end{aligned}
	}
\end{equation}

For histories longer than $k+1$, the definition is irrelevant and can be arbitrary as we shall only consider $(k+1)$-step optimality.
These strategies yield correct over-/under-approximations in the sense of Theorem~\ref{thm:svi-sg-ecfree} below, lifting Theorem~\ref{thm:svi-mc}. 
Formally, this is stated as Lemma~\ref{lem:svi-sg-ecfree} below, with constant bounds $\glb,\gub$,  that we generalize to variable bounds in Theorem~\ref{thm:svi-sg-ecfree}.

\begin{lemma} \label{lem:svi-sg-ecfree}
	Let $\G$ be an SG whose state space is partitioned into $\targets$, $\sinks$ and $\unknown$.  Let $\glb,\gub \in \Reals \cap [0,1]$ satisfy $\glb \leq \val(s) \leq \gub$ for all $s \in \unknown$ and let $\sigma_k \in \straas^{\max}, \tau_k \in \straas^{\min}$ be defined by Eq.~\eqref{eq:strats}.
	Then, for every $k \in \Naturals$ and every state $s \in \states$,
	
	\[
	\probability_s^{\sigma_k,\tau_k} (\reach^{\leq k} \targets) +
		\probability_s^{\sigma_{k},\tau_{k}} (\square^{\leq k} \unknown) \cdot \glb
		\leq
		\val(s)
		\leq
		\probability_s^{\sigma_{k},\tau_{k}} (\reach^{\leq k} \targets) +
		\probability_s^{\sigma_{k}, \tau_{k}} (\square^{\leq k} \unknown) \cdot \gub .
    \]
	
\end{lemma}

\begin{proof}
	
	We prove the upper bound by induction on $k$. For all $s \in \unknown$:
	
		\[\val(s) \le
		\probability_s^{\sigma_k,\tau_k}(\reach^{\le k}\targets)
		+
		\probability_s^{\sigma_k,\tau_k}(\square^{\le k}\unknown)\cdot \gub .
		\]

	\para{Base case ($k=0$):}
	We show
	\[
		\probability_s^{\sigma_0,\tau_0}(\reach^{\le 0}\targets)
		+
		\probability_s^{\sigma_0,\tau_0}(\square^{\le 0}\unknown)\cdot \gub
		\ge \val(s).
	\]
	
	If $s \in \targets$, then $\probability_s^{\sigma_0,\tau_0}(\reach^{\le 0}\targets)=1$.
	If $s \notin \targets$, then $\probability_s^{\sigma_0,\tau_0}(\square^{\le 0}\unknown)=1$.
	Since $\gub \ge \val(s)$ for all $s \in \unknown$, the inequality follows.

	\smallskip

	\para{Induction hypothesis:}
	Assume that for some $k \in \Naturals$ and all $s \in \unknown$,
	\begin{align*}
		\probability_s^{\sigma_k,\tau_k}(\reach^{\le k}\targets)
		+
		\probability_s^{\sigma_k,\tau_k}(\square^{\le k}\unknown)\cdot \gub
		\ge
		\val(s).
	\end{align*}

	\para{Induction step:}
	We prove the claim for $k+1$ by distinguishing cases based on the type of state.

	\noindent \emph{Case 1.} $s \in \maxStates$.
	\begin{align*}
		&\probability_s^{\sigma_{k+1},\tau_{k+1}}(\reach^{\le k+1}\targets)
		+
		\probability_s^{\sigma_{k+1},\tau_{k+1}}(\square^{\le k+1}\unknown)\cdot \gub \\
		&=
		\sum_{s'\in\states}\distribution(s,\sigma_{k+1}(s),s')
		\left(
		\probability_{s'}^{\sigma_k,\tau_k}(\reach^{\le k}\targets)
		+
		\probability_{s'}^{\sigma_k,\tau_k}(\square^{\le k}\unknown)\cdot \gub
		\right)
		\tag{by the definition of $\probability^{\sigma_{k+1},\tau_{k+1}}$} \\
		&=
		\max_{a\in\Av(s)}
		\sum_{s'\in\states}\distribution(s,a,s')
		\left(
		\probability_{s'}^{\sigma_k,\tau_k}(\reach^{\le k}\targets)
		+
		\probability_{s'}^{\sigma_k,\tau_k}(\square^{\le k}\unknown)\cdot \gub
		\right)\hspace{0.25in}
		\tag{by the definition of $\sigma_{k+1}$} \\
		&\ge
		\max_{a\in\Av(s)}
		\sum_{s'\in\states}\distribution(s,a,s')\,\val(s')
		\tag{by the induction hypothesis} \\
		&=
		\val(s)
		\tag{by the Bellman equation for $\val$}
	\end{align*}

	\noindent \emph{Case 2.} $s \in \minStates$.
	\begin{align*}
		&\probability_s^{\sigma_{k+1},\tau_{k+1}}(\reach^{\le k+1}\targets)
		+
		\probability_s^{\sigma_{k+1},\tau_{k+1}}(\square^{\le k+1}\unknown)\cdot \gub \\
		&=
		\sum_{s'\in\states}\distribution(s,\tau_{k+1}(s),s')
		\left(
		\probability_{s'}^{\sigma_k,\tau_k}(\reach^{\le k}\targets)
		+
		\probability_{s'}^{\sigma_k,\tau_k}(\square^{\le k}\unknown)\cdot \gub
		\right)
		\tag{by the definition of $\probability^{\sigma_{k+1},\tau_{k+1}}$}  \\
		&\ge
		\sum_{s'\in\states}\distribution(s,\tau_{k+1}(s),s')\,\val(s')\hspace{0.2in}
		\tag{by the induction hypothesis} \\
		&\ge
		\min_{a\in\Av(s)}
		\sum_{s'\in\states}\distribution(s,a,s')\,\val(s')
		\tag{since $\tau_{k+1}(s)\in\Av(s)$} \\
		&=
		\val(s)
		\tag{by the Bellman equation for $\val$}
	\end{align*}
	
The lower bound follows by a symmetric argument. Hence the claim holds for all $k\in\Naturals$.	
	
%
%
	
\end{proof}

To make the notation more readable, from now on we omit the specific strategies $\sigma_k,\tau_k$ and the objectives when possible.%
\newtarget{def:reach-stay} Assuming the players use strategies $\sigma_k$ and $\tau_k$ as defined by Equation~\eqref{eq:strats}, we use $\svireach_s^k \vcentcolon= \probability_s^{\sigma_{k},\tau_{k}} (\reach^{\leq k} \targets)$ for the probability to reach the targets in $k$ steps from $s$ and $\svistay_s^k \vcentcolon= 	\probability_s^{\sigma_{k}, \tau_{k}} (\square^{\leq k} \unknown)$ for the probability to stay in $\unknown$ for $k$ steps from~$s$.

Using trivial bounds of $\glb=0$ and $\gub=1$ is sufficient
for a correct and convergent algorithm, since (i) Theorem~\ref{thm:svi-sg-ecfree}
proves that correctness of the bounds implies correctness of the approximations
and (ii) termination does not depend on the bounds, but rather on the fact that
the staying probability converges to~0. 

However, good bounds are crucial for practical performance, as they enable tight
extrapolation from the step-bounded information accumulated so far. We now develop the necessary machinery to compute such bounds.

\subsection{Decision values}
\label{sec:decision-value}

While the $k$-step strategies from Section~\ref{sec:svi-strats} yield correct
approximations for fixed bounds, updating the bounds is more subtle.

A natural idea is to update the bounds using the closed form of the geometric
series as described in Section~\ref{sec:prelims:svi}, i.e.\ to set $\glb_k = \min_{s \in \unknown} \frac{\svireach_s^k}{1-\svistay_s^k}$ and dually for $\gub$ with taking the maximum.
However, in systems with non-determinism, i.e.\ MDP or SG, doing this is incorrect, as observed in~\cite{DBLP:conf/cav/QuatmannK18} and concisely illustrated by the following example.

\tikzset{node distance=2.5cm, 
	every state/.style={minimum size=15pt, fill=white, circle, 
		align=center, draw}, 
	chance state/.style={minimum size=3pt, inner sep=0pt, fill=black, circle, 
		align=center, draw},
	max state/.style={minimum size=20pt, fill=white, rectangle, 
		align=center, draw},
	every picture/.style={-stealth},
	brace/.style={decorate,decoration=brace}, semithick}

\begin{figure}
		\centering
		\begin{tikzpicture}[scale=1.2, transform shape]
			\node[min vertex, ] (s) {$s$};
			\node[chance state, left of=s, node distance=1cm](c0){};
			\node[chance state, right of=s, node distance=1cm](c1){};
			\node[goal, above left of=c0, node distance=1.5cm] (t0) {};
			\node[goal, above right of=c1, node distance=1.5cm] (t1) {};
			\node[sink, below left of=c0, node distance=1.5cm] (z0) {};
			\node[sink, below right of=c1, node distance=1.5cm] (z1) {};
			
			\draw (t0) to [out=60,in=100,loop,looseness=5](t0);
			\draw (t1) to [out=100,in=60,loop,looseness=5] (t1);
			\draw (z0) to [out=290,in=250,loop,looseness=5] (z0);
			\draw (z1) to [out=250,in=290,loop,looseness=5] (z1);
			\draw (s) to [out=150,in=210,loop,looseness=5] node[left] {$\alpha$}  node[pos=0.15, anchor=north] {\scriptsize{$0.4$}} (s);
			\path[actionedge]
			(s) edge node[action,above] {\scriptsize{$\beta$}} (c1);
			\path (c0) edge node[left]{\scriptsize{$0.4$}} (t0);
			\path (c0) edge node[left]{\scriptsize{$0.2$}}(z0);
			\path (c1) edge node[right] {\scriptsize{$0.5$}}(t1);
			\path (c1) edge node[right] {\scriptsize{$0.5$}}(z1);
		\end{tikzpicture}
		\caption{A Minimizer MDP with $\unknown=\{s\}$}
		\label{fig:dVal}
	\end{figure}

\begin{example}\label{ex:dVal}
	Always playing action $\alpha$ yields $\frac 2 3$ and always playing $\beta$ yields $0.5$. 
	Thus, picking the action with the smaller value, the Minimizer state $s$ has a value of $0.5$.
	
	When using the SVI approach to compute this, we choose the action according to Equation~\eqref{eq:strats}.
	Since the initial lower bound is $\glb_0 = 0$, our choice intuitively only depends on the 1-step reachability probability, which is 0.4 for action $\alpha$ and 0.5 for action $\beta$. 
	Being a Minimizer state, $s$ prefers $\alpha$ and we have $\tau_1(s)=\alpha$.
	
	Computing $\glb_1$ using the naive method (just using the closed form of the geometric series) yields
	$\glb_1 = \frac{\svireach_s^1}{1-\svistay_s^1} = \frac{0.4}{1-0.4}=\frac 2 3$.
	
	This is incorrect! 
	Using action $\beta$, we get $\val(s)=0.5$, so $\frac 2 3$ is not a lower bound on the value.
	Intuitively, the problem is that action $\alpha$ was chosen based on $\glb_0$, where it is optimal.
	But fully committing to it when computing $\glb_1$ ignores the fact that the better action $\beta$ is available.
	When computing the bound, we have to take the other actions into account and not increase the bound to a number where actually a different action would be better.
	
	\begin{figure}
		\centering
		\includegraphics[scale=0.6]{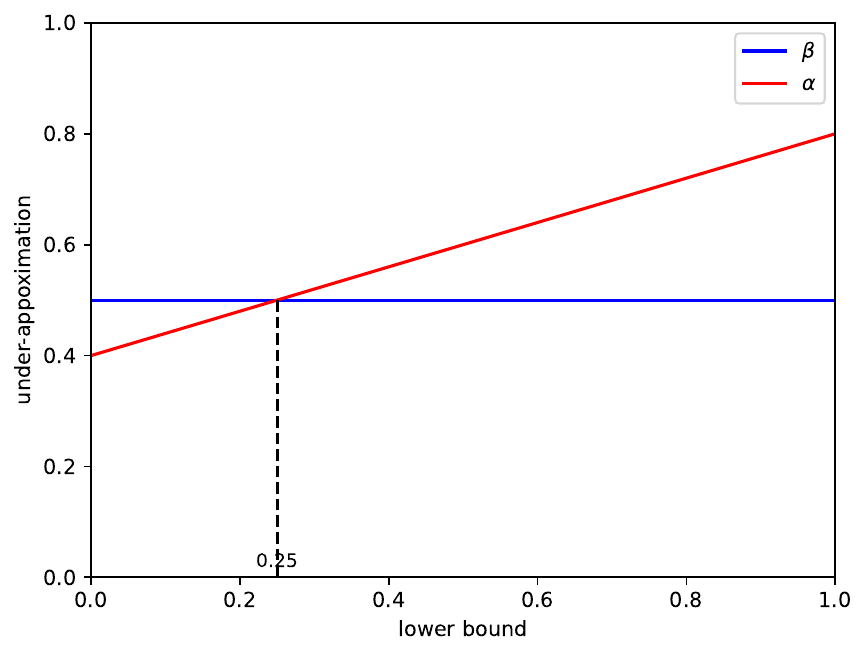}
		\caption{Comparison of actions $\alpha$ and $\beta$ for the MDP in Figure~\ref{fig:dVal}}
		\label{fig:actionsComparison}
	\end{figure}
	
	We illustrate this in Figure~\ref{fig:actionsComparison} where we depict the estimates for action $\alpha$ and $\beta$ depending on the lower bound $\glb$. 
	We see that up until $\glb<0.25$, action $\alpha$ is preferred; for $\glb>0.25$, action $\beta$ yields the smaller estimate. 
	Thus, we must not increase $\glb$ to more than $0.25$, because otherwise the choice of $\alpha$ becomes suboptimal and we lose the correctness guarantees.

\end{example}

Therefore, to preserve correctness, we must restrict the bound so
that the currently chosen action remains optimal. 

Intuitively, each action induces a linear estimate as a function of the
bound, and the preferred action changes at the intersection points of these
lines. 	In~\cite{DBLP:conf/cav/QuatmannK18}, this number was called \emph{decision value}, as the decision of the strategy changes there.

More generally, to ensure correctness, the strategies $\sigma_{k+1},\tau_{k+1}$ must remain optimal not only with respect to $\mathsf{b}_k$, but also with respect to the updated bound $\mathsf{b}_{k+1}$.
Thus, when computing $\mathsf{b}_{k+1}$, we cannot only use the closed form of the geometric series, but we have to pose additional constraints: for every action $\beta$ other than the chosen action $\alpha$, we have to ensure that $\mathsf{b}_{k+1}$ is such that  the chosen action $\alpha$ remains optimal.

Consequently, we want to compute the intersection of the lines in Figure~\ref{fig:actionsComparison} (the decision value) for every pair of actions and enforce that $\mathsf{b}_{k+1}$ cannot cross such a decision value.

Given a state $s$ and a pair of actions $\alpha,\beta$, the decision value is exactly the point where their estimates are equal; the estimate for an action $\alpha$ is defined as $\sum_{s' \in \states} \trans(s,\alpha,s') \cdot (\svireach^k_{s'} + \svistay^k_{s'} \cdot \mathsf{b}_k)$.
Rewriting the equality of the estimates by solving for $\mathsf{b}_k$ yields the decision value:
\begin{align}
	\decval_{\mathsf{b}}^k(s,\alpha,\beta) \eqdef 
	\frac
	{\sum_{s' \in \states} (\trans(s,\beta,s') - \trans(s,\alpha,s'))\cdot\svireach^k_{s'}}
	{\sum_{s' \in \states} (\trans(s,\alpha,s') - \trans(s,\beta,s'))\cdot\svistay^k_{s'}} 
	\defeq \frac{\deltareach(\beta,\alpha)}{\deltastay(\alpha,\beta)}
	\label{eq:def-dval-sab}
\end{align}

Now we can compute the decision value for a pair of actions.
We want to ensure that the lower bound $\glb$ is smaller than the least decision value over all states and actions---the minimum bound where any state would change its decision. Dually, we pick the largest decision value when computing $\gub$. 
However, not every decision value is relevant.
For example, if in Example~\ref{ex:dVal} the current lower bound is already larger than 0.25, we correctly pick action $\beta$, increasing the lower bound will not change our decision and no special precautions have to be taken.
Intuitively, we are on the right of the intersection of the lines in Figure~\ref{fig:actionsComparison}.

Generally, if the slope of the chosen action $\alpha$ is smaller than the slope of the other action $\beta$, i.e.\ $\deltastay(\alpha,\beta)<0$, then the current lower bound is larger than the decision value of these two actions; otherwise $\alpha$ would not have been picked.
Dually, the current upper bound is already smaller than the decision value if $\deltastay(\alpha,\beta)<0$ with chosen action $\alpha$ and other action $\beta$.
Thus, we define the decision value of a state and chosen action which only considers the relevant pairs of actions, i.e.\ only other actions that can become better than the chosen one. 
Note that the conditions for both lower and upper bound are the same as for the minimizer state, the action with the least slope would be the best.

\begin{equation}
	\label{eq:def-dval-sa}
	\boxed{
		\begin{aligned}
			&\decval_{\mathsf{b}}^k(s,\alpha)\eqdef  \opt_{\beta\in\Av(s). \deltastay(\alpha,\beta) > 0} ~ \decval_{\mathsf{b}}^k(s,\alpha,\beta)	\\
			&\displaystyle
			\decval_{\mathsf{b}}^k \eqdef \opt (\decval^{k-1}, \opt_{s \in \unknown} \decval_{\mathsf{b}}^k(s,\alpha)) 
		\end{aligned}
	}
\end{equation}

\subsection{Computing bounds}
\label{sec:no-ec-bounds}
Having identified the decision values that characterize when strategies
change, we now use them to define sound updates of the bounds. The key idea
is to combine the extrapolation obtained from the geometric series with the
restriction that the bound must not cross any relevant decision value.

More precisely, we update the bounds such that (i) they improve based on the
current $k$-step reachability and staying probabilities, and (ii) the chosen
actions remain optimal under the updated bound. For our final definition of the bounds, we utilize a shorthand for the action chosen for state $s$ in the $(k+1)$-st step according to Equation~(\ref{eq:strats}) as $\chosen^{k}(s)\eqdef \arg \opt_{a\in\Av(s)}\sum_{s'\in\states} \trans(s,a,s')\cdot (\svireach_{s'}^{k-1}+\svistay_{s'}^{k-1}\cdot \mathsf{b}_{k-1})$.
Then, as soon as $\svistay^k_s < 1$ for all $s\in\unknown$, we can improve our trivial bounds as follows:

\begin{equation}
	\label{eq:def-bounds}
	\boxed{
		\begin{aligned}
			&\textsc{(Dynamic) Bounds:}\\
			&\mathsf b_k \eqdef  \overline{\opt} \left(\mathsf b_{k-1}, \opt~(\opt_{s \in \unknown}~ \frac{\svireach_s^k}{1-\svistay_s^k}, \decval_{\mathsf{b}}^k) \right)
		\end{aligned}
	}
\end{equation}

Now we can state the correctness of the bounds and of the estimates computed based on the ``$k$-step'' strategies and these dynamically changing bounds:

\begin{restatable}{theorem}{theoremsvisgecfree}
	\label{thm:svi-sg-ecfree} 
	
	Fix an SG $\G$ with its state space partitioned in $\targets$, $\sinks$ and $\unknown$. Let $\glb_k, \gub_k \in \Reals$ be bounds as defined by Equation~\eqref{eq:def-bounds}.
	For $\sigma_k \in  \stratsMax, \tau_k \in \stratsMin$ defined by Equation~\eqref{eq:strats}, 
	the following inequality holds:	
	
	\begin{equation*}\probability_s^{\sigma_k,\tau_k} (\reach^{\leq k} \targets) +
		\probability_s^{\sigma_{k},\tau_{k}} (\square^{\leq k} \unknown) \cdot \glb_k
		\leq
		\val(s)
		\leq
		\probability_s^{\sigma_{k},\tau_{k}} (\reach^{\leq k} \targets) +
		\probability_s^{\sigma_{k}, \tau_{k}} (\square^{\leq k} \unknown) \cdot \gub_k
	\end{equation*}
\end{restatable}

\begin{proof}

	We first prove the upper inequality. The lower is dual and we mention all differences at the end of this proof.	
	
	Fix an arbitrary $k\in\Naturals_0$.
	By Lemma~\ref{lem:global-l-u} (proven below) we have $\val(s)\leq \gub_k$.
	We utilize Inequality~\eqref{eq:maxCorr}, instantiating the arbitrary Minimizer strategy with $\tau_k$ to obtain
	\begin{align}
		\val(s) \leq \sup_{\sigma\in\stratsMax} \left( \probability_{s}^{\sigma,\tau_k} (\reach^{\leq k} \targets) +
		\probability_{s}^{\sigma,\tau_k} (\square^{\leq k} \unknown) \cdot \gub_k \right)
		\label{ineq:corr-above-wo-ec}
	\end{align}
	
	It remains to prove that $\sigma_k$ is among the optimal strategies for the modified objective on the right side of Inequality~\eqref{ineq:corr-above-wo-ec}.
	This part of the proof was omitted in the original paper~\cite{DBLP:conf/cav/QuatmannK18}, just stating that it is necessary but not explicitly proving it.

	\para{Remaining proof goal:} 
	
	For all states $s\in\states$, we have
	\[
	\sup_{\sigma\in\stratsMax} \left( \probability_{s}^{\sigma,\tau_k} (\reach^{\leq k} \targets) +
	\probability_{s}^{\sigma,\tau_k} (\square^{\leq k} \unknown) \cdot \gub_k \right)
	=
	\probability_{s}^{\sigma_k,\tau_k} (\reach^{\leq k} \targets) +
	\probability_{s}^{\sigma_k,\tau_k} (\square^{\leq k} \unknown) \cdot \gub_k
	\]
	
	We proceed by performing an induction on the number of steps remaining in the objective. 
	Before that, we formally define the one step unfolding of the objective. This is necessary for the induction step and instructive for understanding the idea of the proof.
	
	In the following, $s\in\states$ always denotes an arbitrary state and $i\in\Naturals_0$ with $0 < i < k$ denotes a number of steps.
	We highlight that $k$ is an arbitrary but fixed number (thus also fixing $\sigma_k,\tau_k$ and $\glb_k,\gub_k$) and we argue about the step-bounded probabilities used when considering this particular objective.

	\para{Unfolding one step:}
	
	Recall that we use $\pi$ to denote $\sigma$ if it is applied to a state $s\in\maxStates$; and to denote $\tau$ otherwise.
	By standard results, e.g.~\cite[Chapter 4]{Puterman}, we can unfold the definition of the probabilities.
	Formally, for all strategy pairs $(\sigma,\tau)$, we have:
	\[
	\probability_{s}^{\sigma,\tau} (\reach^{\leq i} \targets) = 
	\sum_{s' \in \states} \trans(s,\pi(s),s') \cdot \probability_{s}^{\sigma',\tau'} (\reach^{\leq i-1} \targets)
	\]
	where the updated strategies $\sigma',\tau'$ are defined by prepending $s \pi(s)$ to the history, i.e.\ for every history $\rho$ we have $\pi'(\rho) \eqdef \pi(s \pi(s) \rho)$.
	Note that in particular $\pi_k' = \pi_{k-1}$ because by definition (the second case of the recursion in Eq.~\eqref{eq:strats}) $\pi_k$ mimics the previous strategies.
	
	By analogously unfolding $\probability_{s}^{\sigma,\tau} (\square^{\leq i} \unknown)$ and using distributivity, we obtain that for strategies $\sigma_i,\tau_i$ defined by Eq.~\eqref{eq:strats} the following holds:
	\begin{align*}
		&\probability_{s}^{\sigma_i,\tau_i} (\reach^{\leq i} \targets) +
		\probability_{s}^{\sigma_i,\tau_i} (\square^{\leq i} \unknown) \cdot \gub_k
		\\
		&=
		\sum_{s' \in \states} \trans(s,\pi(s),s') \cdot \left(
		\probability_{s'}^{\sigma_{i-1},\tau_{i-1}} (\reach^{\leq {i-1}} \targets) +
		\probability_{s'}^{\sigma_{i-1},\tau_{i-1}} (\square^{\leq {i-1}} \unknown) \cdot \gub_k
		\right)
	\end{align*}
	
	\para{Base case:}
	\[
	\sup_{\sigma\in\stratsMax} \left( \probability_{s}^{\sigma,\tau_0} (\reach^{\leq 0} \targets) +
	\probability_{s}^{\sigma,\tau_0} (\square^{\leq 0} \unknown) \cdot \gub_k \right)
	=
	\probability_{s}^{\sigma_0,\tau_0} (\reach^{\leq 0} \targets) +
	\probability_{s}^{\sigma_0,\tau_0} (\square^{\leq 0} \unknown) \cdot \gub_k
	\]
	This is true because if there are 0 steps remaining, the choice of strategy is irrelevant.
	Intuitively, this is also why $\pi_0$ picks an arbitrary action in the base case in Eq.~\eqref{eq:strats}.
	
	\para{Induction hypothesis:}
	
	\[
	\sup_{\sigma\in\stratsMax} \left( \probability_{s}^{\sigma,\tau_i} (\reach^{\leq i} \targets) +
	\probability_{s}^{\sigma,\tau_i} (\square^{\leq i} \unknown) \cdot \gub_k \right)
	=
	\probability_{s}^{\sigma_i,\tau_i} (\reach^{\leq i} \targets) +
	\probability_{s}^{\sigma_i,\tau_i} (\square^{\leq i} \unknown) \cdot \gub_k
	\]
	
	Recall that $s\in\states$ always denotes an arbitrary state and $i\in\Naturals$ with $0 < i < k$ denotes a number of steps. We do not care about numbers $j>k$ for our overall goal.
	
	\para{Induction step:}
	
	We make a case distinction on the owner of the arbitrary state $s$.
	We start with the easier case $s\in\minStates$.
	The following chain of equations proves our goal.
	\begin{align*}
		&\sup_{\sigma\in\stratsMax} \left( \probability_{s}^{\sigma,\tau_{i+1}} (\reach^{\leq {i+1}} \targets) +
		\probability_{s}^{\sigma,\tau_{i+1}} (\square^{\leq {i+1}} \unknown) \cdot \gub_k \right)
		\\ 
		&=
		\sum_{s' \in \states} \trans(s,\tau_{i+1}(s),s') \cdot \sup_{\sigma\in\stratsMax} \left(
		\probability_{s'}^{\sigma,\tau_{i}} (\reach^{\leq {i}} \targets) +
		\probability_{s'}^{\sigma,\tau_{i}} (\square^{\leq {i}} \unknown) \cdot \gub_k
		\right)
		\tag{By unfolding one step.}\\
		&=
		\sum_{s' \in \states} \trans(s,\tau_{i+1}(s),s') \cdot 		\probability_{s'}^{\sigma_i,\tau_{i}} (\reach^{\leq {i}} \targets) +
		\probability_{s'}^{\sigma_i,\tau_{i}} (\square^{\leq {i}} \unknown) \cdot \gub_k
		\tag{By induction hypothesis.}\\
		&=
		\probability_{s}^{\sigma_{i+1},\tau_{i+1}} (\reach^{\leq {i+1}} \targets) +
		\probability_{s}^{\sigma_{i+1},\tau_{i+1}} (\square^{\leq {i+1}} \unknown) \cdot \gub_k
		\tag{By reversing the unfolding.}
	\end{align*}
	
	For reversing the unfolding, note that $s\in\minStates$ implies that $\sigma_{i+1}$ is not applied to $s$, so we can just add it to the history.
	This is the added difficulty of the second case where $s\in\maxStates$: proving that the first choice of $\sigma_{i+1}$ is optimal.
	We first provide the chain of equations and then separately provide arguments for the more involved steps.
	
	\begin{align*}
		&\sup_{\sigma\in\stratsMax} \left( \probability_{s}^{\sigma,\tau_{i+1}} (\reach^{\leq {i+1}} \targets) +
		\probability_{s}^{\sigma,\tau_{i+1}} (\square^{\leq {i+1}} \unknown) \cdot \gub_k \right)\\
		&= 
		\sup_{\sigma\in\stratsMax} \left(\sum_{s' \in \states} \trans(s,\sigma(s),s') \cdot \sup_{\sigma'\in\stratsMax} \left(
		\probability_{s'}^{\sigma',\tau_{i}} (\reach^{\leq {i}} \targets) +
		\probability_{s'}^{\sigma',\tau_{i}} (\square^{\leq {i}} \unknown) \cdot \gub_k
		\right)\right)
		\tag{By unfolding one step and Argument 1.}\\
		&=
		\sup_{\sigma\in\stratsMax} \left(\sum_{s' \in \states} \trans(s,\sigma(s),s') \cdot 		\probability_{s'}^{\sigma_i,\tau_{i}} (\reach^{\leq {i}} \targets) +
		\probability_{s'}^{\sigma_i,\tau_{i}} (\square^{\leq {i}} \unknown) \cdot \gub_k
		\right)
		\tag{By induction hypothesis.}\\
		&=
		\sum_{s' \in \states} \trans(s,\sigma_{i+1}(s),s') \cdot 		\probability_{s'}^{\sigma_i,\tau_{i}} (\reach^{\leq {i}} \targets) +
		\probability_{s'}^{\sigma_i,\tau_{i}} (\square^{\leq {i}} \unknown) \cdot \gub_k
		\tag{By Argument 2.}\\
		&=
		\probability_{s}^{\sigma_{i+1},\tau_{i+1}} (\reach^{\leq {i+1}} \targets) +
		\probability_{s}^{\sigma_{i+1},\tau_{i+1}} (\square^{\leq {i+1}} \unknown) \cdot \gub_k
		\tag{By reversing the unfolding.}
	\end{align*}
	
	\textbf{Argument 1:} By unfolding, the strategy for the remaining $i$ steps is $\sigma'(\rho) \eqdef \sigma(s \sigma(s) \rho)$. 
	However, since $\sigma$ is the optimal strategy for $i+1$ steps, we know that $\sigma'$ is the optimal strategy for $i$ steps and can write it as the supremum over all strategies.
	This allows us to immediately apply our induction hypothesis.
	
	\textbf{Argument 2:}
	By definition of the strategies (Eq.~\eqref{eq:strats}), 
	\[\sigma_{i+1}(s) = \argmax_{a\in\Av(s)} \sum_{s' \in S} \distribution(s, a, s') \cdot \Big(\probability^{\sigma_i, \tau_i}_{s'}  (\reach^{\leq i}\targets) + \probability^{\sigma_i, \tau_i}_{s'}  (\stay^{\leq i}\targets) \cdot \gub_i \Big).\]
	The only difference to what we need to prove the equality is that $\sigma_{i+1}$ picks according to $\gub_i$, while the optimal strategy picks according to $\gub_k$.
	This is the reason we have the decision value, and the reason for it being a bound for all iterations to come.
	
	Let $\alpha$ be the action picked by $\sigma_{i+1}$ and $\beta$ the action picked by the optimal strategy. 
	
	By definition of the bounds (Eq.~\eqref{eq:def-bounds}, namely the third item taking the maximum with the previous decision value), we have $\gub_k \geq \decval_{\gub}^i$, and this decision value (by Eq.~\eqref{eq:def-dval-sa}) certainly is greater than $\decval_{\gub}^i(s,\alpha,\beta)$.
	(Side argument: What if $\Delta_{\svistay}(\alpha,\beta)\leq 0$? If it is equal to 0, then by changing the bound the actions still have the same estimate and we know that $\alpha$ is optimal, too. If it is less than 0, decreasing the upper bound will decrease the estimate of $\beta$ by more than $\alpha$, and since $\gub_k\leq\gub_i$ and $\beta$ is optimal, this means $\alpha$ has to be optimal, too, and $\gub_k=\gub_i$.) 
	By definition (Eq.~\eqref{eq:def-dval-sab}), $\decval_{\gub}^i(s,\alpha,\beta)$ is the number where the estimates of $\alpha$ and $\beta$ are equal; for all greater numbers, $\alpha$ has an estimate that is greater or equal to that of $\beta$. Since $\gub_k$ is greater than $\decval_{\gub}^i(s,\alpha,\beta)$, we know that $\alpha$ still is optimal.
	
	Thus, we conclude that picking $\alpha = \sigma_{i+1}(s)$ is also optimal when considering $\gub_k$ and our chain of equations is correct.
	
	\para{Summary and the dual case of the lower bound}
	The proof for the lower bound is dual.
	First, we utilize Inequality~\eqref{eq:minCorr} to reduce our proof goal to showing that $\tau_k$ is an optimal strategy for the step-bounded objective, i.e.\
	\begin{align*}
		\tau_k \in \arginf_{\tau\in\stratsMin} \left( \probability_{s}^{\sigma_k,\tau} (\reach^{\leq k} \targets) +
		\probability_{s}^{\sigma_k,\tau} (\square^{\leq k} \unknown) \cdot \glb_k \right)
	\end{align*}
	Then we perform the same induction on the number of remaining steps, basically always replacing $\max$ with $\min$, $\sup$ with $\inf$ and $\geq$ with $\leq$. 
	We mention that Argument 2 of the induction step utilizes the newly introduced lower decision value; still, the argument remains the same.
	Thus, we conclude that the bounds computed by Algorithm~\ref{alg:svi-sg-high-level} are correct in every iteration.
\end{proof}

\begin{remark}
	The Inequality~\ref{ineq:corr-above-wo-ec} ($\sigma_k$ being an optimal strategy) was not proven in the original paper~\cite{DBLP:conf/cav/QuatmannK18}.
	It only stated its necessity and described how to construct $\sigma_k$.
	We provide the (2-pages long) proof, including the induction that proves the optimality of $\sigma_k$ and utilizes the decision value.
\end{remark}

The following lemma is used at the beginning of the proof of Theorem~\ref{thm:svi-sg-ecfree}.
\begin{restatable}{lemma}{lemmagloballu}
	\label{lem:global-l-u}
	Let $\glb_k$ and $\gub_k$ be defined according to Equation~\eqref{eq:def-bounds}, then $\forall s \in \unknown$ the following holds:
	\begin{enumerate}
		\item $\glb_{k} \leq \val(s)\leq \gub_{k}$
		\item $\glb_{k} \leq \probability_s^{\sigma_{k}, \tau_{k}}(\reach \targets) \leq \gub_{k}$
	\end{enumerate}
\end{restatable}

\para{Proof sketch.} The update of $\gub_k$ decreases the previous bound to a value based on the current strategies. We show that this decrease is safe: either the resulting value remains a valid bound, or the decision value prevents it from becoming too small.
The key idea is to interpret the update of $\gub_k$ as a one-step Bellman-style operator applied to the previous bound $\gub_{k-1}$. Fixing the extremal state $s_{\max}$, this induces an affine map $f(v)=\svireach^k_{s_{\max}}+\svistay^k_{s_{\max}}\cdot v$, whose iterates define a monotonically decreasing sequence converging to the fixpoint $v_\infty = \svireach^k_{s_{\max}}/(1-\svistay^k_{s_{\max}})$, corresponding to the value under the current strategies. Starting from the safe bound $\gub_{k-1}$, one step of this operator preserves the upper-bound property, and thus all iterates remain above the true value. The update of $\gub_k$ then selects a smaller bound by either taking this limit or stopping earlier at the decision value, which prevents the sequence from decreasing beyond what is justified by the current strategies. In both cases, the update maintains the invariant that $\gub_k$ is a valid global upper bound.
The full formal proof is given in Appendix~\ref{sec:appendix}.

\subsection{Algorithm}
\label{sec:algorithm-no-ec}

We summarize the SVI algorithm for SG without ECs in Algorithm~\ref{alg:svi-sg-high-level}.
\begin{algorithm}
	\caption{Sound value iteration for stochastic games without end components}
	\label{alg:svi-sg-high-level}
	\begin{algorithmic}[1]
		\Require SG $\G$ (partitioned into $\unknown,\targets,\sinks$) and precision $\varepsilon>0$
		\Ensure $\val' \colon \states \to \Reals$ such that $\abs{\val'(s)-\val(s)} \leq \varepsilon$ for all $s \in \states$

		\State For $s\in \targets$ and all $k\in\Naturals_0$, $\svireach_s^k \gets 1$ and $\svistay_s^k \gets 0$  \label{line:svi:initStart}
		\State For $s\in \sinks$ and all $k\in\Naturals_0$, $\svireach_s^k \gets 0$ and $\svistay_s^k \gets 0$
		\State For $s\in \unknown$, $\svireach_s^0 \gets 0$ and $\svistay_s^0 \gets 1$ 
		\State $\glb_0\gets0$ and $\gub_0\gets1$ \label{line:svi:initEnd}
		\State $k\gets 0$
		
		\smallskip
		
		\Repeat
		\State Use \textsc{``$k$-step'' Strategies} to choose $\sigma_{k+1}$ and $\tau_{k+1}$---see Equation~\eqref{eq:strats} \label{line:svi:chooseStrats}
		\For{all $s\in\unknown$} 
		\State Compute $\svireach_s^{k+1}$ and $\svistay_s^{k+1}$ by \textsc{Bellman-Update} 
		---see Equation~\eqref{alg:update-no-ec}
		\label{line:svi:updateStepBoundProbs}
		\EndFor
		\If{\textsc{updateGlobalBounds}?---see condition~\eqref{eqn:updateGlobalBounds-no-ec}} 
		\State Compute $\glb_{k+1}$ and $\gub_{k+1}$ by \textsc{(Dynamic) Bounds}---see Equation~\eqref{eq:def-bounds}
		\label{line:svi:updateGlobalBounds-l-u}
		\EndIf
		\State $k \gets k+1$
		\Until{$\svistay^k_s \cdot (\gub_k-\glb_k) < 2\cdot\varepsilon$ for all $s\in\unknown$} \label{line:svi:terminationCond}
		\State	\Return $\svireach^k_s + \svistay^k_s \cdot \frac{\glb_k+\gub_k}{2}$ for all $s \in \unknown$
		
	\end{algorithmic}
\end{algorithm}

We start by initializing the 0-step bounded probabilities and upper and lower bounds using the trivial under- and over-approximations of 0 and 1 (Lines~\ref{line:svi:initStart}-\ref{line:svi:initEnd}).
Note that for targets and sinks, this trivial initialization is done for all step bounds.
Afterwards, we enter the main loop of the algorithm which repeats the following steps:
firstly, as discussed in Section~\ref{sec:svi-strats}, in Line~\ref{line:svi:chooseStrats} we choose the strategies optimizing the $k$-step objective $\probability^{\sigma_k,\tau_k}(\reach^{\leq k} \targets) + \probability^{\sigma_k,\tau_k}(\stay^{\leq k} \unknown)\cdot \mathsf b_k$. 
Note that this only relies on $k$-step values which have been computed in the previous iteration.
Secondly, in Line~\ref{line:svi:updateStepBoundProbs} we compute the $(k+1)$-step bounded probabilities for all the states according to our strategies, again building on the $k$-step values:

\begin{equation}
	\label{alg:update-no-ec}
	\boxed{
		\begin{aligned}
			&\textsc{Bellman-Update:}\\
			&\svireach^{k+1}_s \gets \sum_{s' \in \states}\distribution(s,\pi_{k+1}(s), s') \cdot \svireach^{k}_{s'} ;~~~\svistay^{k+1}_{s} \gets \sum_{s' \in \states}\distribution(s,\pi_{k+1}(s), s') \cdot \svistay^{k}_{s'}
		\end{aligned}
	}
\end{equation}

Next, in Line~\ref{line:svi:updateGlobalBounds-l-u}, we update the global bounds as specified in Eq.~\eqref{eq:def-bounds}, but only under the condition denoted \textsc{`updateGlobalBounds?'} .
The condition intuitively means the we can only update the global bounds when all the states in $\unknown$ have $\svistay$ strictly less than 1, otherwise $\frac{\svireach_s^k}{1-\svistay_s^k}$ would be undefined.
\begin{equation}
	\boxed{
		\begin{aligned}
			\label{eqn:updateGlobalBounds-no-ec}
			\textsc{updateGlobalBounds}?:= \forall s \in \unknown. \svistay^k_s< 1 
		\end{aligned}
	}
\end{equation}

These updates of $\svireach$, $\svistay$ and the bounds are repeated for increasing $k$ until the approximations of lower and upper bounds are close enough (Line~\ref{line:svi:terminationCond}). Intuitively, this amounts to checking whether $\svistay$ is close to 0 or the global lower and upper bound are close to each other. One can also use relative difference by replacing $\svistay^k_{s}$ by $\frac{\svistay^k_{s}}{\svireach^k_{s}+\svistay^k_{s} \cdot \gub}$ which could be useful in various applications. We give the following theorem:

\begin{restatable}{theorem}{theoremsvinoec}
	\label{thm:svi:no-ec}
Fix an SG $\G$ that contains no ECs, with its state space partitioned in $\targets$, $\sinks$ and $\unknown$, and fix a precision $\varepsilon>0$. 
Algorithm~\ref{alg:svi-sg-high-level} is correct and terminates with an $\varepsilon$-close approximation to the value $\val'$, i.e. for all $s \in \unknown, |\val'(s)- \val(s)| \leq \varepsilon$.
\end{restatable}
\begin{proof}
	
	\para{Correctness.} 
	In Lemma~\ref{lem:svi:inv-1}, we prove that Algorithm~\ref{alg:svi-sg-high-level} computes the stepbounds $\probability_s^{\sigma_k,\tau_k} (\reach^{\leq k} \targets)$ and $\probability_s^{\sigma_k, \tau_k}(\square^{\leq k} \unknown)$, as well as bounds $\glb_k$ and $\gub_k$ that are aligned with Lemma~\ref{lem:global-l-u}. 
	Then, correctness directly follows from Theorem~\ref{thm:svi-sg-ecfree}.

	\para{Termination.} 
	Any play of an MDP or SG, independent of the strategies, eventually remains inside a single MEC with probability one~\cite{dA97a}.   Under the assumption that SG  does not contain an EC in $\unknown$, we have $\lim_{k \to \infty} \probability_s^{\sigma_k, \tau_k} (\square^{\leq k} \unknown) = 0$. 
	This implies that 
	$\val(s)  =\lim_{k \to \infty} \probability_s^{\sigma_k,\tau_k} (\reach^{\leq k} \targets)$.
\end{proof}

	\begin{remark}
		The crucial difficulty for the correctness is that $\svireach^k_s$ and $\svistay^k_s$ indeed correspond to the correct step-bounded probabilities for the modified objective, proven in Lemma~\ref{lem:svi:inv-1} below.
        The core difficulty arises from the dynamic updating of the bounds $\glb$ and $\gub$.
        For static bounds, the correctness proof is a lot simpler, and in fact exactly what we showed in Lemma~\ref{lem:svi-sg-ecfree} above.
	\end{remark}

\begin{restatable}{lemma}{sviInvriantOne}
	\label{lem:svi:inv-1}
	After executing $k \in \Naturals_0$ iterations of Algorithm~\ref{alg:svi-sg-high-level}, the following holds for all $s \in \unknown$:
	\begin{enumerate}
		\item $\svireach^k_s=\probability_s^{\sigma_k,\tau_k} (\reach^{\leq k} \targets)$
		\item $\svistay^k_s=\probability_s^{\sigma_k,\tau_k} (\stay^{\leq k} \unknown)$
	\end{enumerate}
\end{restatable}

\begin{proof}
	We argue by induction on $k$, with a case distinction on the type of states. 
	Throughout we assume that $\G$ has no end components and that $\pi_k$ is the ``$k$-step'' strategy from Equation~\eqref{eq:strats}; 
	we write $(\sigma_k,\tau_k)$ for the strategies induced by~$\pi_k$.
	
	\smallskip
	\paragraph{Base case:}
	For $k=0$, $\pi_0$ (and hence $\sigma_0,\tau_0$) is arbitrary. By initialization we have
	\begin{align*}
		\svireach^0_s &= 1, &&\forall s \in \targets, &
		\svireach^0_s &= 0, &&\forall s \in \states \setminus \targets,\\
		\svistay^0_s  &= 1, &&\forall s \in \unknown, &
		\svistay^0_s  &= 0, &&\forall s \in \states \setminus \unknown.
	\end{align*}
	Hence, for all $s$,
	\[
	\svireach^0_s = \probability_s^{\sigma_0,\tau_0} (\reach^{\leq 0} \targets),
	\qquad
	\svistay^0_s  = \probability_s^{\sigma_0,\tau_0} (\stay^{\leq 0} \unknown).
	\]
	
	\smallskip
	\paragraph{Induction hypothesis:}
	Fix $k \ge 0$ and assume that for all $s \in \unknown$,
	\begin{align*}
		\svireach^k_s &= \probability_s^{\sigma_k,\tau_k} (\reach^{\leq k} \targets),\\
		\svistay^k_s  &= \probability_s^{\sigma_k,\tau_k} (\stay^{\leq k} \unknown),
	\end{align*}
	
	\smallskip
	\paragraph{Inductive step:}
	Consider iteration $k+1$. 
	The strategy $\pi_{k+1}$ is defined by Equation~\eqref{eq:strats}: for a state $s$,
	\[
	\pi_{k+1}(s) \in \arg\opt_{a\in\Av(s)} 
	\sum_{s' \in \states} \distribution(s, a, s') 
	\Big(\probability^{\sigma_k, \tau_k}_{s'}  (\reach^{\leq k}\targets) 
	+ \probability^{\sigma_k, \tau_k}_{s'}  (\stay^{\leq k}\unknown) \cdot \mathsf b_k \Big),
	\]
	where $\opt$ is either $\max$ (for Maximizer states) or $\min$ (for Minimizer states), and $\mathsf b_k$ is the global upper bound $\gub_k$ defined earlier.
	Let $\alpha = \pi_{k+1}(s)$ be the action chosen in $s$.
	
	The values $\svireach^{k+1}$ and $\svistay^{k+1}$ are then updated by the Bellman step
	(see Equation~\eqref{alg:update-no-ec} and Line~\ref{line:svi:updateStepBoundProbs}):
	\begin{equation*}
		\svireach^{k+1}_s \gets \sum_{s' \in \states}\distribution(s,\pi_{k+1}(s), s') \cdot \svireach^{k}_{s'} ,
		\qquad
		\svistay^{k+1}_{s} \gets \sum_{s' \in \states}\distribution(s,\pi_{k+1}(s), s') \cdot \svistay^{k}_{s'} .
	\end{equation*}
	
	Using the induction hypothesis, we obtain for all $s \in \unknown$:
	\begin{align*}
		\svireach^{k+1}_s
		&= \sum_{s' \in \states} \distribution(s,\alpha, s') \cdot \svireach^{k}_{s'} \tag{Line~\ref{line:svi:updateStepBoundProbs} of Algorithm~\ref{alg:svi-sg-high-level} and Equation~\eqref{alg:update-no-ec}.}\\
		&= \sum_{s' \in \states} \distribution(s,\alpha, s') 
		\cdot \probability^{\sigma_k, \tau_k}_{s'}  (\reach^{\leq k}\targets) 
		\tag{by the induction hypothesis}\\
		&= \probability^{\sigma_{k+1},\tau_{k+1}}_s (\reach^{\leq k+1}\targets),
	\end{align*}
	where the last equality uses that we first choose $\alpha$ in $s$ and then follow $(\sigma_k,\tau_k)$
	for another $k$ steps. The argument for $\svistay^{k+1}_s$ is analogous and yields
	\[
	\svistay^{k+1}_s
	= \probability^{\sigma_{k+1},\tau_{k+1}}_s (\stay^{\leq k+1}\unknown).
	\]

	It remains to justify that the action $\alpha = \pi_{k+1}(s)$ is chosen optimally w.r.t.\ the current values.
	Define for each $k$, state $s$, and action $a \in \Av(s)$:
	\[
	f_k(s,a) \coloneqq
	\sum_{s' \in \states} \distribution(s, a, s')
	\Big(\probability^{\sigma_k, \tau_k}_{s'}  (\reach^{\leq k}\targets) 
	+ \probability^{\sigma_k, \tau_k}_{s'}  (\stay^{\leq k}\unknown) \cdot \mathsf b_k \Big).
	\]
	By Equation~\eqref{eq:strats}, for every state $s$,
	\[
	\pi_{k+1}(s) \in \arg\opt_{a\in\Av(s)} f_k(s,a),
	\]
	i.e.\ $\alpha$ maximizes $f_k(s,\cdot)$ in Maximizer states and minimizes it in Minimizer states. 
	Thus, in either case, $\alpha$ is an action that is optimal with respect to $f_k$. This completes the proof.
\end{proof}

%% file: 4_svi-with-ec.tex
\section{Extending sound value iteration to systems with end components}
\label{sec:algorithm-with-ec}

We now extend SVI to SGs with end components (ECs). 
We first show that Algorithm~\ref{alg:svi-sg-high-level} does not converge in the presence of ECs. 
We then show that natural extensions of the \emph{deflate} approach for BVI~\cite{KKKW18} do not resolve this issue for SVI. 
This leads us to two new notions, namely the \emph{best-exit set (BES)} and the \emph{delay} action, which together yield a convergent variant of SVI.

\subsection{Why standard SVI fails in the presence of ECs} 
\label{sec:non-convergence}
\tikzset{node distance=2.5cm, 
	every state/.style={minimum size=15pt, fill=white, circle, 
		align=center, draw}, 
	chance state/.style={minimum size=3pt, inner sep=0pt, fill=black, circle, 
		align=center, draw},
	max state/.style={minimum size=20pt, fill=white, rectangle, 
		align=center, draw},
	every picture/.style={-stealth},
	brace/.style={decorate,decoration=brace}, semithick}

Standard SVI may fail to converge in the presence of end components (ECs) because the termination (Theorem~\ref{thm:svi:no-ec}) crucially relies on the absence of ECs. 
We illustrate this on the simple example in Figure~\ref{fig:example1}, where $\unknown=\{s\}$.
We apply Algorithm~\ref{alg:svi-sg-high-level} to this example.
After initialization, we have $\glb=0$, $\gub=1$, $\svireach_s^k=0$, and $\svistay_s^k=1$. 
For every $k \geq 1$, the algorithm chooses $\sigma(s):=a$, thereby staying inside the EC and leaving the approximations unchanged. 
Hence, the over-approximation remains equal to $1$ and does not converge to the true value $\nicefrac12$.

This shows that standard SVI is insufficient in the presence of ECs. 
The core difficulty is that updates within an EC are circular and therefore non-informative. 
The key element of any solution to the EC issue is to identify where to best \emph{exit} the EC, which provides meaningful updates on the value compared to non-informative updates within the EC. In the following, we examine natural approaches to handling ECs and show why they are insufficient.

\begin{figure}
	\begin{tikzpicture}
		\node[max vertex,initial ] (s) {$s$};
		\node[chance state] at (1,0)(c){};
		\node[goal] at (2.5,0.5) (f) {};
		\node[sink] at (2.5,-0.5) (z) {};
		
		\draw (f) to [out=100,in=60,loop,looseness=5] (f);
		\draw (z) to [out=290,in=250,loop,looseness=5] (z);
		\draw (s) to [out=60,in=120,loop,looseness=5] node[above] {$a$} (s);
		\path[actionedge]
		(s) edge node[action,above] {$b$} (c);
		\path (c) edge node[above] {\scriptsize{$\nicefrac12$}} (f);
		\path (c) edge node[below] {\scriptsize{$\nicefrac12$}}(z);
	\end{tikzpicture}
	\caption{A game with trivial Maximizer EC}
	\label{fig:example1}
\end{figure}

\para{Root of the difficulties: Incompatibility of SVI and deflating in BVI.}
ECs are handled in \emph{BVI} using \emph{deflating}~\cite{KKKW18}, which identifies \emph{simple end components (SECs)}--ECs in which all states have the same value. 
Within an SEC, deflating reduces the local over-approximations of all states to the value of Maximizer's \emph{best exit} (see Definition~\ref{def:bestExit}), since none of these states can possibly get any higher value.

In contrast, SVI does not maintain explicit local over-approximations for each state  which could be reduced (as used in deflating for BVI). 
Instead, it derives them from the $k$-step quantities $\svireach$, $\svistay$, and a global bound $\gub$. 
As a result, deflating cannot be applied directly, and adapting it to SVI becomes non-trivial.

Moreover, since SVI represents values via a step-bounded unfolding, handling ECs requires a finer decomposition than SECs.

We outline the key difficulties below using illustrative examples.
\begin{enumerate}[(1)]
	\item Although SVI maintains an over-approximation $\gub$, it is a global bound shared by all states and thus cannot be used to locally deflate the value of states within a given EC.
	
	\begin{example}
		Consider the SG in Figure~\ref{fig:3stateLocalU}. We have $\unknown = \{s_0, s_1, s_2\}$, with $s_2$ as the initial state. 
		The set $X = \{s_0, s_1\}$ forms a simple end component (SEC). 
		Analyzing the best exit $\bestExit^\Box(X)=\{(s_0,\mathsf{b}), (s_1,\mathsf{b})\}$ shows that the maximal achievable value within $X$ is $0.5$. 
		 However, since $s_2 \notin X$, we can not reduce the upper bound of $s_2$ (because in general $s_2$ may have actions leading to higher values).
		This shows that deflating a global bound is insufficient.
	\end{example}
	
	\begin{figure}[!ht]
		\centering
		\begin{tikzpicture}
			\node[max vertex, ] (s0) {$s_0$};
			\node[max vertex, right of=s0] (s1) {$s_1$};
			\node[max vertex, below right of=s0, node distance=1.75cm] (s2) {$s_2$};
			\node[chance state, left of=s0, node distance=1cm](c0){};
			\node[chance state, right of=s1, node distance=1cm](c1){};
			\node[goal, above left of=s0, node distance=2cm] (t0) {};
			\node[goal, above right of=s1, node distance=2cm] (t1) {};
			\node[sink, below left of=s0, node distance=2cm] (z0) {};
			\node[sink, below right of=s1, node distance=2cm] (z1) {};
			
			\path (s0) edge[bend left=30] node[above] {$a$} (s1);
			\path (s1) edge[bend left=30] node[above] {$a$} (s0);
			
			\path (s2) edge[bend left=10] node[below] {$a$} (s0);
			\path (s2) edge[bend right=10] node[below] {$b$} (s1);
			
			\draw (t0) to [out=110,in=170,loop,looseness=7] (t0);
			\draw (t1) to [out=70,in=10,loop,looseness=7] (t1);
			\draw (z0) to [out=290,in=250,loop,looseness=8] (z0);
			\draw (z1) to [out=250,in=290,loop,looseness=8] (z1);
			\draw (s0) to [out=150,in=210,loop,looseness=5] node[left] {$b$} node[pos=0.15, anchor=north] {\scriptsize{$0.2$}} (s0);
			\draw (s1) to [out=-30,in=30,loop,looseness=5]  node[right] {$b$} node[pos=0.15, anchor=south] {\scriptsize{$0.2$}}(s1);
			\path (c0) edge node[pos=0.6, anchor=east] {\scriptsize{$0.4$}} (t0);
			\path (c0) edge node[pos=0.6, anchor=east] {\scriptsize{$0.4$}} (z0);
			\path (c1) edge  node[pos=0.6, anchor=west] {\scriptsize{$0.4$}}(t1);
			\path (c1) edge  node[pos=0.6, anchor=west] {\scriptsize{$0.4$}} (z1);
		\end{tikzpicture}
		\caption{SG with 3 maximizer nodes, self-loops, and a 2-state SEC}
		\label{fig:3stateLocalU}
	\end{figure}

	This motivates a natural extension where $\svireach$ and $\svistay$ are also updated. 
	However, as we show next, deflating an entire SEC by adjusting these quantities is also problematic.

	\item SVI maintains $\svireach$ and $\svistay$, from which over-approximation is derived.
	Since the same over-approximation can arise from multiple pairs of these values, deflating it is inherently ambiguous. 
	Consequently, when deflating a state, it is unclear to which pair of values to deflate; moreover, there may not even exist any choice that works correctly for the entire EC (or SEC).
	
	\begin{example}
		Consider the SG in Figure~\ref{fig:deflate-example} (left).
		For all states $s \in \unknown=\{s_0,s_1,s_2\}$, Algorithm~\ref{alg:svi-sg-high-level} yields $\svireach_s=0$, $\svistay_s=1$, and $\glb=0$, $\gub=1$, resulting in an over-approximation $\svireach_s+\svistay_s \cdot \gub = 1$.
		
		Based on these approximations, the entire set $\unknown$ is identified as a candidate SEC, although this is not an actual SEC with respect to the true values. Let $T := \unknown$.
		
		The best exit $\bestExit^\Box(T)=\{(s_2,\mathsf{c'})\}$ yields $\svireach=0.6$ and $\svistay=0$, matching the true value of $s_2$.
		
		Deflating all states in $T$ to $0.6$ would give the correct over-approximation. However, this cannot be achieved consistently: setting $\svireach_{s_1}=0.6$ would violate the requirement that $\svireach$ is an under-approximation, since the true reachability probability from $s_1$ is at most $0.4$.
	\end{example}
	
	This motivates a natural alternative where only best-exit states are deflated. 
	However, as we show next, this approach is also insufficient.
	
	\begin{figure}[!ht] 
		\begin{center}
			
			\input{deflate-example1} 
			\input{deflate-example2} 
		\end{center}
		\caption{SG where deflating cannot be easily applied to fix SVI}
		\label{fig:deflate-example} 
	\end{figure}			
	\item Deflating only the best-exit states of a SEC may leave the remaining states unchanged, resulting in non-convergence within the SEC.
	
	\begin{example}
		Consider the SG in Figure~\ref{fig:deflate-example} (right).
		The SEC $\{\initstate, s_1\}$ has the best exit $(s_1,\mathsf{d})$. 
		Deflating the over-approximation of $s_1$ to $0.5$ yields $\svireach_{s_1}=0.5$ and $\svistay_{s_1}=0$.
		
		However, this update must also propagate to $s_0$. 
		Otherwise, due to action $\mathsf{a}$, we retain $\svireach_{s_0}=0$ and $\svistay_{s_0}=1$, resulting in an over-approximation of $1$, which does not converge to the true value $\nicefrac12$.
	\end{example}
    This motivates that we need to recursively find and deflate best exits within ECs. 
    However, as we show next, this is still insufficient.
    \item There can be cyclical depencies between the best exits:
	\begin{example}
		\label{ex:2stateCyclicDependency}
		Consider the SG in Figure~\ref{fig:2stateCyclicDependency}. 
		It contains a MEC $T = \unknown = \{s_0,s_1\}$.
		Both these are Maximizer states and have value $\nicefrac{1}{2}$, achieved by playing at least one of the exiting actions infinitely often.
		Upon initialization, we have: 
		$\svireach^0_{s_0}=0,\svistay^0_{s_0}=1,\svireach^0_{s_1}=0,\svistay^0_{s_1}=1,\glb_0=0, \gub_0=1$.
		Comparing the exits from $T$, for $(s_0,b)$, the approximation $\svireach+\svistay\cdot\gub$ evaluates to $\nicefrac{1}{3}+\nicefrac{1}{3}\cdot1=\nicefrac{2}{3}$, which exceeds the approximation for $(s_1, b)$ given by $0.4+0.2\cdot1=0.6$.  
		Therefore, the best exit of $T$ in the first iteration is $\bestExit^\Box(T)=\{(s_0, b)\}$. 
		Forcing Maximizer to take this best exit leads to the following updated vectors:
		$\svireach^1_{s_0}=\nicefrac13,\svistay^1_{s_0}=\nicefrac13, \svireach^1_{s_1}=0, \svistay^1_{s_1}=1,
		\glb_1=0, \gub_1=1.$
	
		In the second iteration, taking  $(s_0, b)$ again would decrease  $\svistay_2^{s_0}$ to $\nicefrac13\cdot\nicefrac13=\nicefrac19$ while increasing $\svireach_2^{s_0}$ to $\nicefrac13\cdot 1+\nicefrac13\cdot\nicefrac13=\nicefrac49$, yielding $\nicefrac49+\nicefrac19\cdot 1=\nicefrac59$.
		This is less than one step of $(s_1,b)$ yielding $0.4+0.2\cdot1=0.6$.
		Consequently, updating the best exit results in: 
		$\svireach^2_{s_0}=0,\svistay^2_{s_0}=1,\svireach^2_{s_1}=0.4, \svistay^2_{s_1}=0.2,\glb_2=0 ,\gub_2=1.$
		In the third iteration, $\{(s_0, b)\}$ again is the best exit, which results in updating the vectors to the same values as in the first iteration.
		This oscillation continues indefinitely.
		As a result, the bounds remain unchanged, i.e.\ $\glb_i=0$, $\gub_i=1$  for all the subsequent iterations.
		
		The non-termination arises because the algorithm keeps shifting the probability of remaining undecided entirely between two states (as reflected by $\svistay=1$ for either $s_0$ or $s_1$ in each iteration).  
		To address this, we enforce monotonic progress of $\svireach+\svistay\cdot\gub$ (see Equation~\eqref{eqn:impUB}); if updating a state would break monotonicity, it instead plays a ``delay'' action, see Section~\ref{ssec:new-notions}.
	
	\end{example}
	
	\item Both $\svireach$ and $\svistay$ are defined with respect to a step-count $k$. 
	However, collapsing ECs ignores steps taken within the component, while deflating effectively skips multiple steps by jumping to an exit. 
	In contrast, SVI requires an exact accounting of steps. 
	To address this, we introduce a ``delay'' action in Section~\ref{ssec:new-notions}, which keeps the state unchanged while still counting as a step.
	
\end{enumerate}	
	
	We highlight that difficulties (1), (3), (4) and (5)
	already arise in MDP, since the respective examples only contain Maximizer states.
	Overall, deflating an entire SEC to its best exit harms correctness; and, deflating only the best exiting state $\bestExit_f^\Box$ leads to non-termination. Therefore, handling ECs in SVI requires a different mechanism than SEC-based deflation.

	\begin{figure}
		\centering
		\begin{tikzpicture}
			\node[max vertex, ] (s0) {$s_0$};
			\node[max vertex, right of=s0] (s1) {$s_1$};
			\node[chance state, left of=s0, node distance=1cm](c0){};
			\node[chance state, right of=s1, node distance=1cm](c1){};
			\node[goal, above left of=s0, node distance=2cm] (t0) {};
			\node[goal, above right of=s1, node distance=2cm] (t1) {};
			\node[sink, below left of=s0, node distance=2cm] (z0) {};
			\node[sink, below right of=s1, node distance=2cm] (z1) {};
			
			\path (s0) edge[bend left=30] node[above] {$a$} (s1);
			\path (s1) edge[bend left=30] node[below] {$a$} (s0);
			\draw (t0) to [out=110,in=170,loop,looseness=7] (t0);
			\draw (t1) to [out=70,in=10,loop,looseness=7] (t1);
			\draw (z0) to [out=290,in=250,loop,looseness=8] (z0);
			\draw (z1) to [out=250,in=290,loop,looseness=8] (z1);
			\draw (s0) to [out=150,in=210,loop,looseness=5] node[left] {$b$} node[pos=0.15, anchor=north] {\scriptsize{\nicefrac{1}{3}}} (s0);
			\draw (s1) to [out=-30,in=30,loop,looseness=5]  node[right] {$b$} node[pos=0.15, anchor=south] {\scriptsize{$0.2$}}(s1);
			\path (c0) edge node[pos=0.6, anchor=east] {\scriptsize{\nicefrac{1}{3}}} (t0);
			\path (c0) edge node[pos=0.6, anchor=east] {\scriptsize{\nicefrac{1}{3}}} (z0);
			\path (c1) edge  node[pos=0.6, anchor=west] {\scriptsize{0.4}}(t1);
			\path (c1) edge  node[pos=0.6, anchor=west] {\scriptsize{0.4}} (z1);
		\end{tikzpicture}
		\caption{Game graph with two maximizer nodes and self-loops}
		\label{fig:2stateCyclicDependency}
	\end{figure}

\subsection{Our solution}

We resolve ECs using a new approach that updates the value vectors consisting of two key ideas: 1) we introduce a ``recursive variant'' of deflating that identifies best exits by recursively cutting ECs, thereby all breaking cyclic dependencies; and 2) we define a ``delay condition'' that determines whether to apply a standard update or to remain in the current state by playing a ``delay'' action, enforcing monotonicity and progress of upper bounds.

Importantly, our approach remains within the spirit of SVI, as it does not require maintaining local upper bounds for individual states. 
Our solution can be seen as a refinement of Algorithm~\ref{alg:svi-sg-high-level}, where the computation of strategies and updates is adapted using the procedures introduced below.

\subsubsection{\textbf{Best exit set (BES)}}\label{ssec:bes}
This section examines the hierarchical structure of ECs and their respective exits. 
The procedure \textsc{BestExitSet} recursively identifies the best exits for each sub-EC, resulting in a comprehensive set of the different best exits---capturing relevant state-action pairs that serve as exits, each corresponding to some EC.
We remark that the set of best exits is empty for ECs that cannot be exited at all.

\begin{algorithm}[!ht]
	\caption{Compute best exit set for the sound value iteration} 
	\label{alg:bestexitset-high-level}
	\begin{algorithmic}[1]
		\Require an SG $\G$, over-approximation $f\geq \val$, an EC $\mec$
		\Ensure Set of correct exiting state-action pairs of ECs in $\mec$ wrt $f$
		\Procedure{BestExitSet}{$\G, f, \mec$}
		\State BES $\gets \emptyset$ \Comment{Initialization}

		\If {$\bestExit^\Box_f(\mec)$ is empty} \label{line:BES-hl:BestExitEmpty}\Comment{Base case}
		\State Move all states of $\mec$ into $\sinks$ and exclude them from $\unknown$ \label{line:BES-hl:removeTrapECs}
		\State \Return BES \label{line:BES-hl:returnBaseCase}

		\Else \Comment{Recursive step: {$\bestExit^\Box_f(\mec)$ is non-empty}}
		\State Include all best-exits of $\mec$ to BES \label{line:BES-hl:insertBE}
		\State Compute MECs of the induced graph after removing all states of the best-exits \label{line:BES-hl:MECsInInducedGraph}
		\State Recursively compute the BES for each MEC obtained in the previous step and include the resulting state-action pairs in BES \label{line:BES-hl:callBESOnAllMECs}
		\EndIf
		\State \Return BES {\Comment{Set of correct exits \emph{wrt} $f$ for all EC within $\mec$}}\label{line:BES-hl:return2}
		\EndProcedure
	\end{algorithmic}
\end{algorithm}

Algorithm~\ref{alg:bestexitset-high-level} defines the procedure \textsc{BestExitSet}, which computes the \emph{best exit set (BES)} for a given EC $\mec$ with respect to an over-approximation $f \geq \val$. The BES contains all relevant exiting state-action pairs under the current approximation.
If $\bestExit^\Box_f(\mec)$ is empty (Lines~\ref{line:BES-hl:BestExitEmpty}--\ref{line:BES-hl:returnBaseCase}), then the EC has no Maximizer exits.  This implies that if the play is in any of the states in this EC, then\phantomsection\newtarget{def:trap}{the Minimizer can force the play to remain there forever; we call such ECs \emph{trap ECs}.}
We denote the set of all such ECs as \trap. 
All states of $\mec$ are therefore added to the sink set $\sinks$ and removed from the unknown set $\unknown$ (Line~\ref{line:BES-hl:removeTrapECs}).
Otherwise, the best exits of $\mec$ are added to BES (Line~\ref{line:BES-hl:insertBE}). 
After removing these states, the procedure recursively processes all MECs of the induced graph (Lines~\ref{line:BES-hl:MECsInInducedGraph}--\ref{line:BES-hl:callBESOnAllMECs}). Finally, Line~\ref{line:BES-hl:return2} returns the set of best exits for all relevant sub-ECs of $\mec$.

\medspace
The recursive nature of the procedure leverages the inductive, attractor-like structure of MECs, providing a systematic decomposition that identifies all relevant exits while avoiding cyclic dependencies.
Importantly, when the best exit of a MEC is identified removed, this often eliminates multiple sub-ECs
simultaneously. This hierarchical dependency inherently reduces the effective number of ECs that must
be considered in subsequent recursive calls: It is linear in the number of states in the MEC despite the
potential exponential total number of ECs. 

Next, we establish a fundamental property of best exits, which shows that the value of every state in an EC is bounded by the value of the best exit computed under the current over-approximation. 
This property is then used to prove soundness and completeness of the procedure \textsc{BestExitSet}.

\begin{lemma}
	\label{lem:bestExit}
	Given an SG $\G$, an EC $\mec \subseteq \unknown$, and any $f:\states \to [0,1]$ with $f \geq \val$,
	if $\bestExit^{\Box}_{\val}(\mec)=(s_{\val},a_{\val})$ and $\bestExit^{\Box}_{f}(\mec)=(s_f,a_f)$, then for all $s \in \mec$,
	\[
	\val(s) \leq f(s_f,a_f).
	\]
\end{lemma}

\begin{proof}
	We prove the claim in three steps:
	\begin{enumerate}
		\item $\forall s \in \mec:\ \val(s) \leq \val(s_{\val},a_{\val})$,
		\item $\val(s_{\val},a_{\val}) \leq f(s_{\val},a_{\val})$,
		\item $f(s_{\val},a_{\val}) \leq f(s_f,a_f)$.
	\end{enumerate}

	\begin{enumerate}

	\item We prove that for every $s \in \mec$, one has $\val(s) \leq \val(s_{\val},a_{\val})$, by distinguishing the type of state.
	
	\medspace

	\noindent \emph{Case 1: $s \in \mec_{\circ}$.}
	At a Minimizer state, the value is
	\[
	\val(s)=\min_{a \in \Av(s)} \val(s,a).
	\]
	Hence, for every action $a \in \Av(s)$,
	\[
	\val(s) \leq \val(s,a).
	\]
	%
	%
	If $(s,a)$ exits $\mec$, then by the definition of $\bestExit^{\Box}_{\val}(\mec)$,
	\[
	\val(s,a) \leq \val(s_{\val},a_{\val}).
	\]
	If $(s,a)$ stays inside $\mec$, then continuing from successors of $(s,a)$ cannot yield a value larger than the best possible exit value from $\mec$; otherwise there would be some exit from $\mec$ with value strictly larger than $\val(s_{\val},a_{\val})$, contradicting the definition of $\bestExit^{\Box}_{\val}(\mec)$. Thus again
	\[
	\val(s,a) \leq \val(s_{\val},a_{\val}).
	\]
	Therefore $\val(s) \leq \val(s_{\val},a_{\val})$.
	
	\medspace
	
	\noindent\emph{Case 2: $s \in \mec_{\Box}$.}
	Since every play that reaches the target with positive probability must eventually leave $\mec$, no strategy starting in $s$ can achieve a value larger than the value of the best exit from $\mec$. Hence,
	\[
	\val(s) \leq \val(s_{\val},a_{\val}).
	\]
	Combining both cases yields
	\[
	\forall s \in \mec:\ \val(s) \leq \val(s_{\val},a_{\val}).
	\]
	
	\item
	This follows directly from $f \geq \val$, hence
	\[
	\val(s_{\val},a_{\val}) \leq f(s_{\val},a_{\val}).
	\]
	
	\item
	Since $(s_f,a_f) \in \bestExit_f^{\Box}(\mec)$, it maximizes $f$ over all exits of $\mec$. Therefore, if
	\[
	f(s_{\val},a_{\val}) > f(s_f,a_f),
	\]
	then $(s_f,a_f) \notin \bestExit_f^{\Box}(\mec)$, a contradiction. Hence
	\[
	f(s_{\val},a_{\val}) \leq f(s_f,a_f).
	\]
	
		\end{enumerate}
	Combining Steps (1), (2), and (3), we obtain
	\[
	\forall s \in \mec:\ \val(s) \leq f(s_f,a_f).
	\]
\end{proof}

Finally, we prove the following property of Algorithm~\ref{alg:bestexitset-high-level}: it is sound insofar that every state-action pair included in the BES has a higher over-approximation than value; and
for every sub-EC, some exit is included in the BES (or it has been merged into the sink states).

\begin{restatable}{lemma}{lemmaBestExitProps}
	\label{lem:bestExitSetProps}
	For every EC $T$ and every $f \geq \val$, Algorithm~\ref{alg:bestexitset-high-level} terminates and returns a set BES with the following properties:
	\begin{enumerate}
		\item[\emph{(I)}] \emph{Soundness of over-approximation:}
		for all $(s,a) \in \text{BES}$, we have
		\[
		f(s,a) \geq \val(s).
		\]
		
		\item[\emph{(II)}] \emph{Completeness of BES:}
		for every EC $T' \subseteq T$,
		\begin{itemize}
			\item if $T' \notin \trap$, then there exists $(s,a) \in \text{BES}$ such that $(s,a)$ exits $T'$;
			\item if $T' \in \trap$, then for all $s \in T'$,
			\(s \in \sinks \land s \notin \unknown.\)
		\end{itemize}
	\end{enumerate}
	Here, $(s,a)$ exits $T'$ means that $\post(s,a) \not\subseteq T'$.
\end{restatable}

\begin{proof}
	We first prove correctness, and then termination.
	
	\para{Correctness.}
	The procedure \textsc{BestExitSet} is first called on an EC $T$ and is then applied recursively to MECs of induced subgraphs obtained after removing states of best exits.
	
	\begin{enumerate}
		\item[\emph{(I)}]
		A pair $(s,a)$ is included in BES only in Line~\ref{line:BES-hl:insertBE}. Thus, there exists an EC $M$ considered during some recursive call such that
		\[
		(s,a) \in \bestExit_f^{\Box}(M).
		\]
		By Lemma~\ref{lem:bestExit}, for every state $s' \in M$ we have
		\[
		\val(s') \leq f(s,a).
		\]
		In particular, since $s \in M$, it follows that
		\[
		f(s,a) \geq \val(s).
		\]
		
		\item[\emph{(II)}]
		Fix an EC $T' \subseteq T$.
		
		If $T' \in \trap$, then by Lines~\ref{line:BES-hl:BestExitEmpty}--\ref{line:BES-hl:returnBaseCase}, all states of $T'$ are moved to $\sinks$ and removed from $\unknown$. Hence the second part holds.
		
		Now assume that $T' \notin \trap$. We show that some pair in BES exits $T'$.
		Consider the first recursive call in which a state of $T'$ is removed from the current subgraph. Such a call exists because either the procedure is eventually applied to $T'$ itself, or some previously chosen best exit removes a state of $T'$ before that happens.
		
		Let $M$ be the EC considered in that call, and let $(s,a) \in \bestExit_f^{\Box}(M)$ be a pair inserted into BES. By construction, $s \in T'$, since this is the first call removing a state of $T'$. Moreover, $(s,a)$ cannot stay inside $T'$, because otherwise removing $s$ would not destroy the presence of $T'$ in the induced subgraph. Hence
		\[
		\post(s,a) \not\subseteq T',
		\]
		that is, $(s,a)$ exits $T'$. Therefore, some exit of $T'$ belongs to BES.
	\end{enumerate}
	
	\para{Termination.}
	In each recursive call, at least one state is removed from the current subgraph before the next recursive calls are made (please refer to Line~\ref{line:BES-hl:MECsInInducedGraph}). Since the state space is finite, only finitely many recursive calls are possible. Therefore, Algorithm~\ref{alg:bestexitset-high-level} terminates.
\end{proof}

Let $f_k(s) := \svireach_s^k + \svistay_s^k \cdot \gub_k$. At each iteration $k$, we compute the overall set $\bestExitSet^k$ as the union of the best-exit sets of all MECs under the current over-approximation. 
\begin{equation}
	\label{eqn:HandleECs}
	\boxed{
		\bestExitSet^k \gets
		\bigcup_{\substack{\mec \subseteq \unknown\\ \mec \text{ is an MEC}}}
		\textsc{BestExitSet}\bigl(\G, \, f_k,\, \mec\bigr)
	}
\end{equation}

Intuitively, $\bestExitSet^k$ collects all currently optimal exits from ECs with respect to $f_k$.

\subsubsection{\textbf{Delays and strategies}}
\label{ssec:new-notions}

The \emph{delay} action is introduced to ensure monotonic progress. 
We refer the reader to Example~\ref{ex:2stateCyclicDependency} for an illustration of the non-termination and cyclic behaviour that may arise without it. 
To determine when a delay action should be played, we first introduce the predicate \textsc{improvedUB}?$(s)$, which checks whether the upper bound at state $s$ improves in the next iteration:
\begin{equation}
	\label{eqn:impUB}
	\boxed{
		\textsc{improvedUB}?(s)
		:=
		\svireach^{k+1}_s + \svistay^{k+1}_s \cdot \gub_k
		\le
		\svireach^k_s + \svistay^k_s \cdot \gub_k
	}
\end{equation}

The following predicate captures when a delay action is played instead of performing the standard update:
\begin{equation}
	\label{eq:delayaction}
	\boxed{
		\textsc{delayAction}?(s)
		:=
		s \in \unknown_{\Box} \land \lnot \textsc{improvedUB}?(s)
	}
\end{equation}

Intuitively, if updating a Maximizer state $s$ would not improve its upper bound, then the algorithm plays a delay action, denoted by $d$, instead of performing the standard update. 
In that case, the approximations at $s$ remain unchanged from iteration $k$ to iteration $k+1$, as specified by the case labeled ``new part for handling ECs'' in Equation~\eqref{alg:update-ec}.

\begin{equation}
	\label{alg:update-ec}
	\boxed{
		\begin{aligned}
			&\textsc{Bellman-Update ($\mathcal{B}$):}\\
			&\mbox{if~}{\textsc{delayAction?}(s)} & \triangleright \mbox{ new part for handling ECs}\\ 
			&\hspace{20pt}\svireach^{k+1}_s \gets \svireach^{k}_{s};~~~\svistay^{k+1}_s \gets \svistay^{k}_{s}\\
			& \mbox{else} & \triangleright \mbox{ standard update: same as before}\\
			&\hspace{20pt}\svireach^{k+1}_s \gets \sum_{s' \in \states}\distribution(s,\pi_{k+1}(s), s') \cdot \svireach^{k}_{s'};~~~&\svistay^{k+1}_{s} \gets \sum_{s' \in \states}\distribution(s,\pi_{k+1}(s), s') \cdot \svistay^{k}_{s'}\\			
		\end{aligned}
	}
\end{equation}

Finally, we redefine the strategies $\sigma_k$ to account for the two new cases of best exits and delay actions, while $\tau_k$ remains unchanged. The strategy gives priority to delay actions and best exits over standard action selection. In particular, delay actions prevent non-improving updates, while best exits enforce progress by leaving end components.

\begin{equation}
	\label{eqn:newstrat}
	\boxed{
		\begin{aligned}
			&\textsc{``$k$-step'' Maximizer's contracting strategies:}\\
			&\sigma_0(\rho) \in \Av(\rho_0),\\
			&\sigma_{k+1}(\rho) \gets
			\begin{cases}
				\sigma_k(\rho') 
				& \text{if } \rho = sa\rho',\\[1mm]
				
				d 
				& \text{if } \rho = s \land \textsc{delayAction?}(s),\\[1mm]
				
				b 
				& \text{if } \rho = s \land (s,b)\in \bestExitSet^k,\\[1mm]
				
				\displaystyle\argmax_{a\in\Av(s)}
				\begin{aligned}[t]
					&\sum_{s' \in \states} \distribution(s,a,s') \cdot \Big( \probability^{\sigma_k,\tau_k}_{s'}(\reach^{\leq k}\targets) \\
					&\qquad + \probability^{\sigma_k,\tau_k}_{s'}(\stay^{\leq k}\unknown)\cdot \gub_k
					\Big)
				\end{aligned}
				& \text{otherwise.}
			\end{cases}
		\end{aligned}
	}
\end{equation}

The combination of best-exit set and delay actions yields a well-defined update scheme that avoids circular dependencies within ECs while preserving monotonic progress. In the next section, we establish correctness and convergence of the resulting algorithm.

\subsection{\textbf{SVI for SGs with ECs}}

We first establish correctness of the approximations and strategies in the presence of ECs.
Compared to the EC-free case, the proof only needs to handle two additional cases, namely best-exit actions and delay actions.

\begin{restatable}{theorem}{theoremSVIGenSGEC}
	\label{thm:svi-gen-sg-ec} 

	Let $\G$ be an SG with state space partitioned into $\targets$, $\sinks$, and $\unknown$.  Let $\glb_k, \gub_k \in \Reals$ be bounds as defined in Equation~\eqref{eq:def-bounds}, i.e. i.e., for all $k \geq 0$ and all $s \in \unknown$, $\glb_k \leq \val(s) \leq \gub_k$.
	Let $\sigma_k \in \stratsMax$ and $\tau_k \in \stratsMin$ be defined as in Equations~\eqref{eqn:newstrat} and~\eqref{eq:strats}, respectively. 
	Then, for all $s \in \unknown$, the following inequality holds:
	\begin{align*}
	\probability_s^{\sigma_k,\tau_k} (\reach^{\leq k} \targets) +
	\probability_s^{\sigma_k,\tau_k} (\stay^{\leq k} \unknown) \cdot \glb_k
	\;\leq\;
	\val(s)
	\;\leq\;
	\probability_s^{\sigma_k,\tau_k} (\reach^{\leq k} \targets) +
	\probability_s^{\sigma_k,\tau_k} (\stay^{\leq k} \unknown) \cdot \gub_k.
\end{align*}

\end{restatable}

\begin{proof}
		We follow the proof of Theorem~\ref{thm:svi-sg-ecfree}. As in the EC-free case, we first prove the upper inequality by reducing it to the optimality of the Maximizer strategy for the modified step-bounded objective
		\[
		\probability_s^{\sigma,\tau_k}(\reach^{\le k}\targets)
		+
		\probability_s^{\sigma,\tau_k}(\stay^{\le k}\unknown)\cdot \gub_k.
		\]
		The proof of Theorem~\ref{thm:svi-sg-ecfree} already covers all standard states. Hence it remains to show that the two additional cases in Equation~\eqref{eqn:newstrat}, namely best-exit actions and delay actions, preserve this optimality argument.
	
	\smallskip
	
	\noindent\emph{Case 1: best-exit action.}
	Assume that $\sigma_{k+1}(s)=b$ because $(s,b)\in\bestExitSet^{k+1}$. 
	By construction of $\bestExitSet^{k+1}$ and Lemma~\ref{lem:bestExitSetProps}, the value of every state in the corresponding EC is bounded by the value of the chosen exit under the current over-approximation. Hence, replacing the standard maximizing choice by the enforced action $b$ does not decrease the value of the modified step-bounded objective, and therefore remains optimal with respect to the one-step unfolding of the step-bounded objective.
	
	\smallskip
	
	\noindent\emph{Case 2: delay action.}
	Assume that $\sigma_{k+1}(s)=d$. By Equation~\eqref{alg:update-ec}, the update keeps $\svireach_s$ and $\svistay_s$ unchanged, and therefore also preserves the value of the modified step-bounded objective at state $s$. Thus, the delay action does not violate the inductive optimality argument.
	
	With these two additional Maximizer cases accounted for, the remainder of the proof is identical to Theorem~\ref{thm:svi-sg-ecfree}. Hence the upper inequality follows.
	
	The lower inequality is obtained dually, as in Theorem~\ref{thm:svi-sg-ecfree}, with the same two additional cases handled analogously.
\end{proof}

\para{Algorithm.} Now we describe how Algorithm~\ref{alg:svi-sg-high-level} is instantiated to handle ECs.
First, we use the Maximizer strategy defined in Equation~\eqref{eqn:newstrat}.
Second, we use the modified Bellman update from Equation~\eqref{alg:update-ec}.
Finally, we update the global bounds only under the following condition:

\begin{equation}
	\label{eqn:updateGlobalBounds}
	\boxed{
		\begin{aligned}
			\textsc{updateGlobalBounds}?
			\;:=\;
			\forall s \in \unknown:\;
			\svistay_s^k < 1 \;\land\; \lnot \textsc{delayAction?}(s)
		\end{aligned}
	}
\end{equation}

\begin{remark}
	Introducing delay actions changes the semantics of 
	$\svireach_s^k$ and 
	$\svistay_s^k$: they now correspond to optimal reachability and stay probabilities within \emph{at most} $k$ effective steps. This does not affect correctness, as delay actions only stutter without changing the reachability value.
	Further, the global bounds are updated only when \textsc{updateGlobalBounds}? holds. 
	In particular, if a delay action is played, the bounds are not updated. 
\end{remark}

We now establish the key invariants of the instantiated algorithm.
In particular, we show that the computed quantities $\svireach_s^k$ and $\svistay_s^k$ correspond to the step-bounded probabilities induced by the strategies $\sigma_k$ and $\tau_k$, and that the global bounds remain sound throughout the iterations.

\begin{lemma}
	\label{lem:svi:inv-1-ec}
	After executing $k \in \Naturals_0$ iterations of Algortihm~\ref{alg:svi-sg-high-level} instantiated with the sub-procedures in Section~\ref{sec:algorithm-with-ec} (Maximizer strategy~\eqref{eqn:newstrat}, new \textsc{Bellman-update}~\eqref{alg:update-ec} and \textsc{isUpdateGlobalBounds?}~\eqref{eqn:updateGlobalBounds}), the following holds for all $s \in \unknown$: 
	
	\begin{enumerate}
		\item $\svireach^k_s=\probability_s^{\sigma_k,\tau_k} (\reach^{\leq k} \targets)$
		\item $\svistay^k_s=\probability_s^{\sigma_k,\tau_k} (\stay^{\leq k} \unknown)$ 
		\item $\glb_k \leq \val(s) \leq \gub_k$ and  $\glb_k \leq \probability_s^{\sigma_k,\tau_k} (\reach \targets) \leq \gub_k$
	\end{enumerate}
\end{lemma}

\begin{proof}
	We prove the lemma by induction on $k$.
	
	The cases for $s \in \minStates$ and for $s \in \maxStates$ where actions are selected as in Equation~\eqref{eq:strats} are identical to Lemma~\ref{lem:svi:inv-1}. 
	It remains to consider the two additional Maximizer cases.
	
	\smallskip
	\noindent\emph{Delay action.}
	If $\textsc{delayAction?}(s)$ holds, then the update sets
	\[
	\svireach_s^{k+1} = \svireach_s^k
	\qquad\text{and}\qquad
	\svistay_s^{k+1} = \svistay_s^k.
	\]
	Thus, the values correspond exactly to the probabilities under $\sigma_{k+1},\tau_{k+1}$, where the play may stutter at $s$. This preserves both (1) and (2).
	
	\smallskip
	\noindent\emph{Best-exit action.}
	If $(s,a) \in \bestExitSet^k$ and $\lnot\textsc{delayAction?}(s)$, then $\sigma_k(s)=a$. 
	Hence the Bellman update follows the same one-step unfolding as in the EC-free case, and according to the current values of the approximations, playing the action $\alpha$ in state $s$ is the best choice that leaves the EC (By the definition of $\bestExit_f^\Box$). This establishes (1) and (2).
	
	\smallskip
	Finally, property (3) follows directly from Lemma~\ref{lem:global-l-u}, since the updates to $\glb_k$ and $\gub_k$ are conservative.
\end{proof}

A crucial ingredient for proving termination is the monotonicity of the over-approximation, which we establish next.

\begin{restatable}{lemma}{lemmamonotonicU}
	\label{lem:monotonicU}
Fix $\gub, \glb$ such that $\glb \leq \val(s) \leq \gub$ for all $s \in \unknown$. Algorithm~\ref{alg:svi-sg-high-level} instantiated with Maximizer's stategy using Equation~\eqref{eqn:newstrat}, and \textsc{updateGlobalBounds?} in Equation~\eqref{eqn:updateGlobalBounds} computes a monotonically non-increasing sequence of $\probability_s^{\sigma_{k},\tau_{k}} (\reach^{\leq k} \targets) +
\probability_s^{\sigma_{k}, \tau_{k}} (\stay^{\leq k} \unknown) \cdot \gub$,
i.e.\
\begin{align*}
	\probability_s^{\sigma_{k+1},\tau_{k+1}} (\reach^{\leq {k+1}} \targets) +
	\probability_s^{\sigma_{k+1},\tau_{k+1}} (\stay^{\leq {k+1}} \unknown) \cdot \gub \leq
	\probability_s^{\sigma_k,\tau_k} (\reach^{\leq k} \targets) +
	\probability_s^{\sigma_{k},\tau_{k}} (\stay^{\leq k} \unknown) \cdot \gub
\end{align*}
	
\end{restatable}

\begin{proof}
	We prove it by case split and use induction.

	\para{Base case:} For $k=0$ and every $s \in \unknown$,
	\[
	\probability_s^{\sigma_{0},\tau_{0}} (\reach^{\leq {0}} \targets) = 0,
	\qquad
	\probability_s^{\sigma_{0},\tau_{0}} (\stay^{\leq {0}} \unknown) = 1,
	\]
	and hence
	\[
	\probability_s^{\sigma_{0},\tau_{0}} (\reach^{\leq {0}} \targets) +
	\probability_s^{\sigma_{0},\tau_{0}} (\stay^{\leq {0}} \unknown) \cdot \gub
	= \gub.
	\]
	
	\para{Induction hypothesis:}
	For some $k \ge 0$ and for all $s \in \unknown$,
	\begin{align*}
		\probability_s^{\sigma_{k+1},\tau_{k+1}} (\reach^{\leq {k+1}} \targets) +
		\probability_s^{\sigma_{k+1},\tau_{k+1}} (\stay^{\leq {k+1}} \unknown) \cdot \gub \leq
		\probability_s^{\sigma_{k},\tau_{k}} (\reach^{\leq {k}} \targets) +
		\probability_s^{\sigma_{k},\tau_{k}} (\stay^{\leq {k}} \unknown) \cdot \gub
	\end{align*}

	\para{Induction step:}
	We show that for all $s \in \unknown$,
	\begin{align*}
		&\probability_s^{\sigma_{k+2},\tau_{k+2}} (\reach^{\leq {k+2}} \targets) +
		\probability_s^{\sigma_{k+2},\tau_{k+2}} (\stay^{\leq {k+2}} \unknown) \cdot \gub \\
		\leq\;&
		\probability_s^{\sigma_{k+1},\tau_{k+1}} (\reach^{\leq {k+1}} \targets) +
		\probability_s^{\sigma_{k+1},\tau_{k+1}} (\stay^{\leq {k+1}} \unknown) \cdot \gub.
	\end{align*}
	
	\noindent\emph{Case 1:} The action chosen at $s$ does not change from iteration $k+1$ to iteration $k+2$.

	\begin{align*}
		&\probability_s^{\sigma_{k+2},\tau_{k+2}} (\reach^{\leq {k+2}} \targets) +
		\probability_s^{\sigma_{k+2},\tau_{k+2}} (\stay^{\leq {k+2}} \unknown) \cdot \gub\\
		=\;& \sum_{s' \in \states}\distribution(s,a,s') \cdot
		\left[
		\probability_{s'}^{\sigma_{k+1},\tau_{k+1}} (\reach^{\leq k+1} \targets) +
		\probability_{s'}^{\sigma_{k+1},\tau_{k+1}} (\stay^{\leq k+1} \unknown) \cdot \gub
		\right]
		\tag{by unfolding one step}\\
		\hspace{0.5in}\leq\;& \sum_{s' \in \states}\distribution(s,a,s') \cdot
		\left[
		\probability_{s'}^{\sigma_{k},\tau_{k}} (\reach^{\leq k} \targets) +
		\probability_{s'}^{\sigma_{k},\tau_{k}} (\stay^{\leq k} \unknown) \cdot \gub
		\right]
		\tag{by the induction hypothesis}\\
		=\;& \probability_s^{\sigma_{k+1},\tau_{k+1}} (\reach^{\leq k+1} \targets) +
		\probability_s^{\sigma_{k+1},\tau_{k+1}} (\stay^{\leq k+1} \unknown) \cdot \gub
		\tag{by folding back}
	\end{align*}

	\noindent\emph{Case 2:} The action chosen at $s$ changes from iteration $k+1$ to iteration $k+2$.

	We distinguish according to the owner of $s$ and the rule used by the strategy update.
	\begin{enumerate}
		\item $s \in \maxStates$: We consider the cases according to Maximizer's strategy in Equation~\eqref{eqn:newstrat}.

	\begin{enumerate}
		\item \textsc{delayAction?} is played in $s$.
		In this case, the strategy either selects the delay action or switches to another action
		only if \textsc{ImprovedUB}$(s)?$ holds.
		Hence the value cannot increase, i.e.
		\begin{align*}
			&\probability_s^{\sigma_{k+2},\tau_{k+2}} (\reach^{\leq {k+2}} \targets) +
			\probability_s^{\sigma_{k+2},\tau_{k+2}} (\stay^{\leq {k+2}} \unknown) \cdot \gub \\
			\le\;&
			\probability_s^{\sigma_{k+1},\tau_{k+1}} (\reach^{\leq {k+1}} \targets) +
			\probability_s^{\sigma_{k+1},\tau_{k+1}} (\stay^{\leq {k+1}} \unknown) \cdot \gub.
		\end{align*}
		
		\item $(s,b) \in \bestExitSet^{k+2}$.
		In this case, $s$ belongs to an EC and the strategy selects a best-exit action.
		We distinguish whether the previous action at $s$ was already a best-exit action.
			\begin{enumerate}
				\item $(s, a_{old}) \in \bestExitSet^{k+1}$.
				\begin{itemize}
					\item $\textsc{delayAction}?(s)$.
					In this case, the delay action induces a self-loop with probability $1$, hence
					\begin{align*}
						&\probability_s^{\sigma_{k+2},\tau_{k+2}} (\reach^{\leq {k+2}} \targets) +
						\probability_s^{\sigma_{k+2},\tau_{k+2}} (\stay^{\leq {k+2}} \unknown) \cdot \gub\\
						=\;&
						\probability_s^{\sigma_{k+1},\tau_{k+1}} (\reach^{\leq {k+1}} \targets) +
						\probability_s^{\sigma_{k+1},\tau_{k+1}} (\stay^{\leq {k+1}} \unknown) \cdot \gub.
					\end{align*}

					\item $\lnot\textsc{delayAction}?(s)$.
					Hence the strategy selects the best-exit action $b$ at iteration $k+2$.
					By definition of the best-exit choice, the action $b$ is optimal among the exit actions with respect to the current objective.
					\begin{align*}
						&\probability_s^{\sigma_{k+2},\tau_{k+2}} (\reach^{\leq {k+2}} \targets) +
						\probability_s^{\sigma_{k+2},\tau_{k+2}} (\stay^{\leq {k+2}} \unknown) \cdot \gub\\
						=\;& \sum_{s' \in \states}\distribution(s,b,s') \cdot
						\left[
						\probability_{s'}^{\sigma_{k+1},\tau_{k+1}} (\reach^{\leq k+1} \targets) +
						\probability_{s'}^{\sigma_{k+1},\tau_{k+1}} (\stay^{\leq k+1} \unknown) \cdot \gub
						\right]\\
						\hspace{1.0in}\leq\;& \sum_{s' \in \states}\distribution(s,b,s') \cdot
						\left[
						\probability_{s'}^{\sigma_{k},\tau_{k}} (\reach^{\leq k} \targets) +
						\probability_{s'}^{\sigma_{k},\tau_{k}} (\stay^{\leq k} \unknown) \cdot \gub
						\right]
						\tag{by the induction hypothesis}\\
						\leq\;& \sum_{s' \in \states}\distribution(s,a_{old},s') \cdot
						\left[
						\probability_{s'}^{\sigma_{k},\tau_{k}} (\reach^{\leq k} \targets) +
						\probability_{s'}^{\sigma_{k},\tau_{k}} (\stay^{\leq k} \unknown) \cdot \gub
						\right]
						\tag{since $a_{old}$ was optimal at iteration $k+1$}\\
						=\;& \probability_s^{\sigma_{k+1},\tau_{k+1}} (\reach^{\leq k+1} \targets) +
						\probability_s^{\sigma_{k+1},\tau_{k+1}} (\stay^{\leq k+1} \unknown) \cdot \gub.
					\end{align*}
					
				\end{itemize}

				\item $(s, a_{old}) \not\in \bestExitSet^{k+1}$.
				\begin{itemize}
					\item $a_{old} \neq d$.
					In this case, $a_{old}$ was chosen at iteration $k+1$ to maximize
					\(
					\sum_{s' \in \states}\distribution(s,a,s') \cdot
					\left(
					\probability_{s'}^{\sigma_k,\tau_k} (\reach^{\leq k} \targets) +
					\probability_{s'}^{\sigma_k,\tau_k} (\stay^{\leq k} \unknown)\cdot \gub
					\right).
					\)
					
					\begin{align*}
						&\probability_s^{\sigma_{k+2},\tau_{k+2}} (\reach^{\leq {k+2}} \targets) +
						\probability_s^{\sigma_{k+2},\tau_{k+2}} (\stay^{\leq {k+2}} \unknown) \cdot \gub\\
						=\;& \sum_{s' \in \states}\distribution(s,b,s') \cdot
						\left[
						\probability_{s'}^{\sigma_{k+1},\tau_{k+1}} (\reach^{\leq k+1} \targets) +
						\probability_{s'}^{\sigma_{k+1},\tau_{k+1}} (\stay^{\leq k+1} \unknown) \cdot \gub
						\right]\\
						\hspace{1.0in}\leq\;& \sum_{s' \in \states}\distribution(s,b,s') \cdot
						\left[
						\probability_{s'}^{\sigma_{k},\tau_{k}} (\reach^{\leq k} \targets) +
						\probability_{s'}^{\sigma_{k},\tau_{k}} (\stay^{\leq k} \unknown) \cdot \gub
						\right]
						\tag{by the induction hypothesis}\\
						\leq\;& \sum_{s' \in \states}\distribution(s,a_{old},s') \cdot
						\left[
						\probability_{s'}^{\sigma_{k},\tau_{k}} (\reach^{\leq k} \targets) +
						\probability_{s'}^{\sigma_{k},\tau_{k}} (\stay^{\leq k} \unknown) \cdot \gub
						\right]
						\tag{since $a_{old}$ was chosen optimally at iteration $k+1$}\\
						=\;& \probability_s^{\sigma_{k+1},\tau_{k+1}} (\reach^{\leq k+1} \targets) +
						\probability_s^{\sigma_{k+1},\tau_{k+1}} (\stay^{\leq k+1} \unknown) \cdot \gub.
					\end{align*}

					\item $a_{old} = d$.
					Since the action changes from delay to a non-delay action at iteration $k+2$,
					the switch is performed only if \textsc{ImprovedUB}$(s)?$ holds.
					Hence the newly chosen action cannot increase
					\(
					\probability_s^{\sigma_{k+2},\tau_{k+2}} (\reach^{\leq {k+2}} \targets) +
					\probability_s^{\sigma_{k+2},\tau_{k+2}} (\stay^{\leq {k+2}} \unknown) \cdot \gub,
					\)
					and therefore
					\begin{align*}
						&\probability_s^{\sigma_{k+2},\tau_{k+2}} (\reach^{\leq {k+2}} \targets) +
						\probability_s^{\sigma_{k+2},\tau_{k+2}} (\stay^{\leq {k+2}} \unknown) \cdot \gub \\
						\le\;&
						\probability_s^{\sigma_{k+1},\tau_{k+1}} (\reach^{\leq {k+1}} \targets) +
						\probability_s^{\sigma_{k+1},\tau_{k+1}} (\stay^{\leq {k+1}} \unknown) \cdot \gub.
					\end{align*}
				\end{itemize}
			\end{enumerate}

			\item The action at $s$ is chosen according to the standard SVI update rule.
			
			If $s$ belongs to an EC, then this case is already covered by the previous cases.
			Otherwise, the selected action maximizes
			\[
			\sum_{s' \in \states}\distribution(s,a,s') \cdot
			\left(
			\probability_{s'}^{\sigma_{k+1},\tau_{k+1}} (\reach^{\leq k+1} \targets) +
			\probability_{s'}^{\sigma_{k+1},\tau_{k+1}} (\stay^{\leq k+1} \unknown)\cdot \gub
			\right).
			\]
			By the induction hypothesis, each action value at horizon $k+1$ is at most its value
			at horizon $k$. Hence the maximum cannot increase, and therefore
			\begin{align*}
				&\probability_s^{\sigma_{k+2},\tau_{k+2}} (\reach^{\leq {k+2}} \targets) +
				\probability_s^{\sigma_{k+2},\tau_{k+2}} (\stay^{\leq {k+2}} \unknown) \cdot \gub \\
				\le\;&
				\probability_s^{\sigma_{k+1},\tau_{k+1}} (\reach^{\leq {k+1}} \targets) +
				\probability_s^{\sigma_{k+1},\tau_{k+1}} (\stay^{\leq {k+1}} \unknown) \cdot \gub.
			\end{align*}
			
		\end{enumerate}
		\item $s \in \minStates$.
		The action is chosen to minimize the objective.
		Hence taking the minimum cannot increase the value, and therefore
		\begin{align*}
			&\probability_s^{\sigma_{k+2},\tau_{k+2}} (\reach^{\leq {k+2}} \targets) +
			\probability_s^{\sigma_{k+2},\tau_{k+2}} (\stay^{\leq {k+2}} \unknown) \cdot \gub \\
			\le\;&
			\probability_s^{\sigma_{k+1},\tau_{k+1}} (\reach^{\leq {k+1}} \targets) +
			\probability_s^{\sigma_{k+1},\tau_{k+1}} (\stay^{\leq {k+1}} \unknown) \cdot \gub.
		\end{align*}
	\end{enumerate}
\end{proof}

\begin{lemma} \label{lem:u-fixpoint}
	(Upper bound converges to a fixpoint). Let \(\ub_i(s) = \svireach_s^i + \svistay_s^i \cdot \gub \). The limit of repeatedly applying the modified Bellman operator \( \mathcal{B} \) to the initial upper bound \( \ub_0 \) exists and is a fixpoint of the operator \( \mathcal{B} \), i.e.,
	
	\[
	\lim_{i \to \infty} \mathcal{B}^i \ub_0 = \mathcal{B} \left( \lim_{i \to \infty} \mathcal{B}^i \ub_0 \right).
	\]
\end{lemma}

\begin{proof}
	We aim to show that repeatedly applying the modified Bellman operator \( \mathcal{B} \) to the initial upper bound \( \ub_0 \) converges to a fixpoint, i.e.,
	
	\[
	\lim_{i \to \infty} \mathcal{B}^i \ub_0 = \mathcal{B} \left( \lim_{i \to \infty} \mathcal{B}^i \ub_0 \right).
	\]
	
	\para{Step 1: Monotonicity of the modified Bellman operator.}
	
	The modified Bellman operator \( \mathcal{B} \) inherits the monotonicity property of standard Bellman operators. Please refer to Lemma~\ref{lem:monotonicU} for a proof of monotonic sequence of upperbounds $\ub_i$. Formally, if \( \ub_1 \leq \ub_2 \), then:
	
	\[
	\mathcal{B}(\ub_1) \leq \mathcal{B}(\ub_2).
	\]
	
	In this case, starting from an initial over-approximation \( \ub_0 \), the sequence of iterates \( \{\mathcal{B}^i \ub_0\}_{i=0}^\infty \) is monotonically non-increasing:
	
	\[
	\ub_0 \geq \mathcal{B} \ub_0 \geq \mathcal{B}^2 \ub_0 \geq \cdots
	\]
	
	Since this sequence is bounded in the interval \( [\val(s), 1] \), it converges to a limit \( \ub^* \), where:
	
	\[
	\ub^* = \lim_{i \to \infty} \mathcal{B}^i \ub_0.
	\]
	
	\para{Step 2: Continuity of the modified Bellman operator.}
	
	We now address the potential sources of discontinuity in the modified Bellman operator \( \mathcal{B} \), specifically related to the choices of strategy \( \sigma_{k+1} \).
	
	\textit{Case (a): Strategy chooses the best exit.}
	
	When the strategy \( \sigma_{k+1} \) chooses the best exit action \( b \) (as defined in Eq.~\eqref{eqn:newstrat}), there is no discontinuity. This is because the value function at the current state now depends on the values outside the EC, and these values evolve continuously. Hence, the update made by \( \mathcal{B} \) respects the continuity of the Bellman update.
	
	\textit{Case (b): Strategy chooses the delay action.}
	
	If the strategy \( \sigma_{k+1} \) chooses the delay action \( d \), then the estimate \( \ub \) remains unchanged. Since there is no modification to the estimate, the operator \( \mathcal{B} \) does not introduce any discontinuity. The Bellman operator in this case remains stable, as it does not alter the values.
	
	Thus, in both cases, the modified Bellman operator preserves continuity. This continuity ensures that for any convergent sequence \( \{\ub_i\} \), we have:
	
	\[
	\lim_{i \to \infty} \mathcal{B}(\ub_i) = \mathcal{B} \left( \lim_{i \to \infty} \ub_i \right).
	\]
	
	Applying this to the sequence \( \{\mathcal{B}^i \ub_0\} \), we get:
	
	\[
	\lim_{i \to \infty} \mathcal{B}^{i+1} \ub_0 = \mathcal{B} \left( \lim_{i \to \infty} \mathcal{B}^i \ub_0 \right).
	\]
	
	Since the limit of \( \{\mathcal{B}^i \ub_0\} \) is \( \ub^* \), this equation becomes:
	
	\[
	\ub^* = \mathcal{B}(\ub^*),
	\]
	
	which shows that \( \ub^* \) is a fixpoint of the modified Bellman operator.

	\para{Step 3: Convergence to the greatest fixpoint.}

	The sequence \( \{\mathcal{B}^i \ub_0\} \) is non-increasing and bounded below. Since we are starting from an initial upper bound \( \ub_0 \), repeated applications of the modified Bellman operator converge to the greatest fixpoint \( \ub^* \), i.e., the largest value satisfying:
	\[
	\ub^* = \mathcal{B}(\ub^*).
	\]
	This occurs because \( \mathcal{B} \) monotonically reduces the approximations until the sequence eventually reaches the greatest fixpoint, at which further applications of \( \mathcal{B} \) leave the value unchanged. Therefore, the limit \( \ub^* \) is the greatest fixpoint of the modified Bellman operator.

	\para{Summary:} 1) The sequence \( \{\mathcal{B}^i \ub_0\} \) is monotonic and bounded below.
	2) The modified Bellman operator \( \mathcal{B} \) is continuous, even with the choice of strategies, and the sequence converges to a fixpoint.
	3) The limit of the sequence is the greatest fixpoint \( \ub^* \).
	
	Thus, we conclude that:
	
	\[
	\lim_{i \to \infty} \mathcal{B}^i \ub_0 = \mathcal{B} \left( \lim_{i \to \infty} \mathcal{B}^i \ub_0 \right),
	\]
	
	which completes the proof.
\end{proof}

With the necessary technical results in place, we can now state the main theorem.
\begin{theorem}
	Fix an SG $\G$ with its state space partitioned in $\targets$, $\sinks$ and $\unknown$ and precision $\varepsilon>0$. 
	Algorithm~\ref{alg:svi-sg-high-level} with sub-procedures in Section~\ref{sec:algorithm-with-ec} (Maximizer strategy~\eqref{eqn:newstrat}, new \textsc{Bellman-update}~\eqref{alg:update-ec} and \textsc{isUpdateGlobalBounds?}~\eqref{eqn:updateGlobalBounds}) is correct and terminates with an $\varepsilon$-close approximation to the value $\val'$, i.e.\ \[\forall s \in \unknown: |\val'(s)- \val(s)| \leq \varepsilon.\]
\end{theorem}
\begin{proof}
	\noindent\textbf{Correctness.} Using Lemma~\ref{lem:svi:inv-1-ec}, we establish that at iteration \(k\),
	\(\svireach_s^k\) and \(\svistay_s^k\) coincide with
	\(\probability^{\sigma_k,\tau_k}_s(\reach^{\le k}\targets)\) and
	\(\probability^{\sigma_k,\tau_k}_s(\stay^{\le k}\unknown)\), respectively.
	Moreover, the same lemma guarantees that the global bounds \(\glb_k\) and
	\(\gub_k\) remain sound.
Hence the assumptions of Theorem~\ref{thm:svi-gen-sg-ec} are satisfied, and correctness follows.

	\smallskip 
	\noindent\textbf{Termination.}
	We give a proof by contradiction of the algorithm termination even in the presence of ECs. We first present a high level idea of the proof in five steps and then present the complete proof. 
	\begin{enumerate}
		\item Firstly, using Lemma~\ref{lem:monotonicU} and the property that $\gub_{k+1} \leq \gub_{k}$, we show that the function $\ub_k$, defined as $\forall s \in \states: \ub_k(s) \eqdef \svireach^k_s + \svistay^k_s \cdot \gub_k$  is well defined and has a fixpoint, i.e. $ \ub^*:=\lim_{k \to \infty}\ub_k$ (Lemma~\ref{lem:u-fixpoint}). 
		\item Secondly, we assume for contradiction that Algorithm~\ref{alg:svi-sg-high-level} does not terminate even when the fixpoint in Step~1 is reached. Using this assumption, we define a subsystem $X$ of $\G$, where the difference between $\ub^*$ and $\val$ is maximum, and show the existence of an EC within $X$.
		\item Next, we show that applying another iteration of Algorithm~\ref{alg:svi-sg-high-level} leads to a contradiction to Step~2 by using Lemma~\ref{lem:bestExit}. In particular, we refute the assumption that $\ub^*>\val$.
		Thus, we can conclude $\ub^*=\val$.
		
		\item Finally, using $\ub^*=\val$ and a similar contradiction argument, we prove that in the limit $\forall s \in \unknown.~ \svistay^k_s=0$. Therefore, the function $\lb_k$, defined as $\forall s \in \states: \lb_k(s) = \svireach^k_s + \svistay^k_s \cdot \glb_k$  equals to $\svireach^k_s$, which implies that under and over-approximations converges to the same fixpoint.
	\end{enumerate}
	
	\noindent Next, we provide the complete termination proof.

    \begin{enumerate}
        \item The sequence \(\svireach_s^k + \svistay_s^k \cdot \gub\) is monotonically non-increasing by Lemma~\ref{lem:monotonicU}, i.e.,
	\begin{equation} \label{eq:monotonic}
		\svireach_s^{k+1} + \svistay_s^{k+1} \cdot \gub \leq \svireach_s^{k} + \svistay_s^{k} \cdot \gub.
	\end{equation}

	Now, consider the sequence \(\ub_k(s) \coloneqq \svireach_s^k + \svistay_s^k \cdot \gub_k\). We have
	
	\begin{align*}
		\ub_{k+1}(s) = \svireach_s^{k+1} + \svistay_s^{k+1} \cdot \gub_{k+1} &\leq \svireach_s^{k+1} + \svistay_s^{k+1} \cdot \gub_k \tag{Since  \(\gub_{k+1} \leq \gub_k\)} \\
		&\leq \svireach_s^k + \svistay_s^k \cdot \gub_k \tag{Using Equation~\eqref{eq:monotonic}}\\
		&= \ub_k(s) \tag{By the definition of \(\ub_k(s)\)}\\ 
		\ub_{k+1}(s) &\leq \ub_k(s)
	\end{align*}
	
	Therefore, the sequence \(\ub_k(s)\) is monotonically non-increasing.

	Furthermore, $\ub_k(s)$  bounded in the interval $[\val(s), 1]$. Therefore, $\ub_k(s)$ converges to a limit as $k \to \infty$. We denote this limit by $\ub^*(s)$, i.e. $\ub^*(s) \eqdef \lim_{k \to \infty}\ub_k(s)$ is well defined. Also, by Lemma~\ref{lem:u-fixpoint}, $\ub^*(s)$ is a fixpoint.
    \item We give a proof by contradiction that $\ub^* = \val$. 
	
	Let us assume that $\exists s : \ub^*(s) - \val (s)>0$. With this assumption we define that there exists a set $X$ of states where the difference between the value $\val$ and the fixpoint $\ub^*$ is maximum. 	
	Formally, if $\Delta (s) = \ub^*(s) - \val(s)$ and $m = \max_{s\in\states} \Delta(s)$, then $X$ is defined as follows.
	\begin{align*}
		X =  \{s~|~\ub^*(s) - \val(s) = m \}
	\end{align*}

	Now, we argue that all the actions that have successors outside the set $X$ have lower value. We overload the notation of $f(s)$ by $f(s,a)$ (e.g. $\val(s)$ by $\val(s,a)$ and $\svistay_s$ by $\svistay(s,a)$). Formally, we give the following claim:
	
	\begin{claim}
		\label{cl:exitingActionLower}
		\begin{align*}
			\forall s \in X~\forall a \in \Av(s): \Big(\post(s,a) \not\subseteq X \Rightarrow	\Delta(s,a)  < \Delta(s)\Big)
		\end{align*}
	\end{claim}
	
	The following chain of equations prove Claim~\ref{cl:exitingActionLower}. 
	\begin{align*}
		\Delta(s,a) &= \ub^*(s,a) - \val(s,a) \tag{By definition of $\Delta$.}\\
		& = \Sigma_{s' \in \states} \distribution(s,a,s') \cdot \Big(\ub^*(s') - \val(s')\Big)\\
		& = \Sigma_{s' \in \states} \distribution(s,a,s') \cdot \Delta(s') \tag{By definition.}\\
		& < \Delta(s) \tag{ $ \exists\  \hat{s}\in \post(s,a): \hat{s} \notin X$, therefore, $\delta(\hat{s}) < m$}
	\end{align*}

	As $\ub^*$ is a fix-point, we have the following.

	\begin{align}
		&\forall s \in \states, \exists a \in \Av(s): \ub^*(s) = \ub^*(s,a) \notag \\
		\Rightarrow &\forall s \in X, \exists a \in \Av(s): \ub^*(s) = \ub^*(s,a) \notag \\
		\Rightarrow & \forall s \in X, \exists a: \post(s,a) \subseteq X \tag{By Claim~\ref{cl:exitingActionLower}.}
	\end{align}

	Last implication shows that there exists a pair of strategies $\pi$ of Maximizer and Minimizer, such that $X$ is closed under probabilistic transitions in the Markov chain induced by $\pi$.
	
	Now, we need to define the notion of \emph{bottom MEC} in $X$ which we adapted from~\cite{KKKW18}. In particular, it is an EC in $X$ such that  every successor of an action leaving the bottom MEC reaches a state outside of $X$ with positive probability under all pairs of policies. Using similar arguments as in Theorem 1 of~\cite{KKKW18}, we can show that there exists an EC $\mec$ that is a bottom MEC in $X$.
	\item 	As $\mec$ is a bottom MEC in $X$. There are two cases possible.
	\begin{enumerate}
		\item For all EC $Z$ such that $Z \supseteq \mec$ with $\bestExit_{\ub}^{\Box}(Z)\neq\emptyset$. This implies that $\exists s \in \mec: (s,a) \in \bestExitSet \land \post(s,a) \not \subseteq X$ (because $\mec$ is a bottom MEC in X).

		We know that $\forall s \in \unknown$, the actions in Algorithm~\ref{alg:svi-sg-high-level} are chosen according to Equations~\eqref{eqn:newstrat} and~\eqref{eq:strats}.

		This further implies that in the next iteration $\ub(s)$ will strictly decrease, which is a contradiction to the assumption that we have reached a fixpoint.
		
		\item There exits an EC $X$ such that $X \supseteq \mec$ with $\bestExit_{\ub}^{\Box}(X) = \emptyset$, i.e. $X$ is \trap. This implies that all the states of $X$ are added to sinks $\sinks$ (Line~\ref{line:BES-hl:removeTrapECs} of Algorithm~\ref{alg:bestexitset-high-level}). This implies that for all states s in $\sinks$ we $\svireach_s=0$ and $\svistay_s=0$. Which implies that $\ub^*(s)=0$ This is contradition to the assumption that $m>0$.
	\end{enumerate}
	
	This proves that the over-approximations converge to the value $\val$.
	
	\medskip 
    \item We now show that the under-approximations converge to $\val$.
	
	Since we have already established that $\ub^* = \val$, it suffices to show that
	\[
	\lim_{k \to \infty} \svistay_s^k = 0
	\qquad \text{for all } s \in \states.
	\]
	
	Assume, for the sake of contradiction, that there exists a state $p$ such that
	\[
	\limsup_{k \to \infty} \svistay_p^k > 0.
	\]
	
	Let
	\[
	m := \sup_{s \in \states} \limsup_{k \to \infty} \svistay_s^k > 0,
	\]
	
	and define
	\[
	Z := \left\{ s \in \states \;\middle|\; \limsup_{k \to \infty} \svistay_s^k = m \right\}.
	\]
	
	By definition of $Z$, for every $s \in Z$ and every action $a \in \Av(s)$ such that $\post(s,a) \not\subseteq Z$, we have
	\[
	\sum_{s'} \distribution(s,a,s') \cdot \limsup_{k \to \infty} \svistay_{s'}^k < m.
	\]
	
	Hence, for all $s \in Z$, there exists an action $a \in \Av(s)$ such that $\post(s,a) \subseteq Z$, i.e., $Z$ is closed under the induced dynamics.
	Therefore, $Z$ contains an end component $\mec \subseteq Z$.
	Observe that $\mec$ contains neither target nor sink states, since for such states $\svistay_s^k = 0$ for all $k$, contradicting $m>0$. Hence, for all $s \in \mec$, the reachability value satisfies $\val(s)=0$. 
	
	However, since $\ub^* = \val$, we obtain $\ub^*(s)=0$ for all $s \in \mec$. By definition of $\ub^*$,
	\[
	\ub^*(s) = \lim_{k \to \infty} \bigl(\svireach_s^k + \svistay_s^k \cdot \gub_k\bigr).
	\]
	
	Since $\gub_k \ge 0$ and $\svireach_s^k \ge 0$, this implies
	\[
	\lim_{k \to \infty} \svistay_s^k = 0,
	\]
	
	contradicting the definition of $\mec \subseteq Z$ with $m>0$.
	Thus, $\lim_{k \to \infty} \svistay_s^k = 0$ for all $s \in \states$.
	It follows that
	\[
	\lim_{k \to \infty} \bigl(\svireach_s^k + \svistay_s^k \cdot \glb_k\bigr)
	=
	\lim_{k \to \infty} \svireach_s^k
	=
	\val(s),
	\]
	
	and hence both lower and upper approximations converge to $\val$.

    \end{enumerate}
	
This implies that both lower and upper bound  converges to the $\svireach$ in the limit.
	
\end{proof}

 \subsection{A prototypical example illustrating fast convergence of SVI}
  \label{svi-good-example}
 \begin{figure}
 	\centering
 	\begin{tikzpicture}
 		\node[max vertex] (s) {$q$};
 		\node[chance state, right of=s, node distance=1cm](c1){};
 		\node[goal, right of=c1, node distance=1.5cm] (t1) {};
 		\node[sink, below of=t1, node distance=2cm] (z1) {};
 		\node[min vertex, below of=s, node distance=2cm] (s') {$r$};
 		\node[initial, min vertex, below left of=s, node distance=1.50cm] (s0) {$p$};
 		\draw (t1) to [out=115,in=65,loop,looseness=6] (t1);
 		\draw (z1) to [out=245,in=290,loop,looseness=6] (z1);
 		\draw (s) to [out=25,in=-25,loop,looseness=6] node[left] {\scriptsize{$a$}}  node[pos=0.15, anchor=south]{\scriptsize{$0.98$}} (s);
 		\path (c1) edge[bend left] node[below] {\scriptsize{$0.01$}}(t1);

 		\path (c1) edge[bend left] node[left] {\scriptsize{$0.01$}}(z1);
 		\path (s0) edge[bend left] node[right] {\scriptsize{$a$}}(s);
 		\path (s0) edge[bend right] node[right] {\scriptsize{$b$}}(s');
 		\path (s') edge[bend right] node[above] {\scriptsize{$c$}}(t1);
 		\path (s) edge[bend right] node[above ] {\scriptsize{$b$}}(z1);
 	\end{tikzpicture}
 	\caption{An SG with $\unknown=\{p,q,r\}$}
 	\label{fig:SG-SVI-good}	

  \end{figure}

Consider the SG in Figure~\ref{fig:SG-SVI-good}.
At the Minimizer state $p$, action $a$ is optimal, since choosing $b$ leads to the target with probability $1$.
At the Maximizer state $q$, action $a$ is optimal, as action $b$ leads directly to the sink state.

This example highlights the advantage of SVI in the presence of probabilistic self-loops.
The loop at state $q$ induces a geometric behavior, which BVI resolves only through repeated value propagation.
In contrast, SVI captures this behavior via its step-bounded formulation, allowing it to account for the effect of the loop in a single iteration.

Empirically, BVI requires a large number of iterations (e.g., 685 under a standard stopping criterion) to approximate the value, whereas SVI converges in just two iterations on this example.

%% file: deflate-example1.tex
	\begin{tikzpicture}
		\drawdummy (init) at (0,0) {};
	    \node[min vertex] (p) at (1,0) {$\initstate$};
		\node[max vertex](q) at (3,1.5) {$s_1$};
		\node[max vertex] (r) at (3,-1.5) {$s_2$};
		\drawdummy (mid) at (4.25,1) {};
		\drawdummy (mid2) at (4.25,-1) {};
		\node[goal] (1) at (5.5,1) {};
	    \node[sink] (0) at (5.5,-1)  {};

		\draw[->] (init) to (p);
		\draw[->]  (p) to[bend left] node [midway,anchor=south] {$\mathsf{a}$}(q) ;
		\draw[->]  (q) to [bend left] node [midway,anchor=north] {$\mathsf{b}$} (p);
		\draw[->]  (p) to[bend left] node [midway,anchor=north] {$\mathsf{a'}$}(r) ;
		\draw[->]  (r) to [bend left] node [midway,anchor=north] {$\mathsf{b'}$}(p);
		\draw[-] (q) to node [midway,anchor=south] {$\mathsf{c}$} (mid) ;
		\draw[-] (r) to node [midway,anchor=south] {$\mathsf{c'}$} (mid2) ;
		\draw[->] (mid) to node [above] {$0.6~~~~~~~~~$} (0);
		\draw[->] (mid) to node [above] {$0.4$} (1);
		\draw[->] (mid2) to node [below] {$0.4$} (0);
		\draw[->] (mid2) to node [below] {$0.6~~~~~~~~~$} (1);
		\draw[->]  (0) to[loop right]  node [midway,anchor=west] {$\mathsf{d}$} (0);
		\draw[->]  (1) to [loop right] node [midway,anchor=west] {$\mathsf{e}$} (1) ;
	\end{tikzpicture}

%% file: deflate-example2.tex
	\begin{tikzpicture}
		\drawdummy (init) at (0,0) {};
		\node[max vertex] (p) at (1,0) {$\initstate$};
		\node[max vertex] (q) at (3,0) {$s_1$};
		\drawdummy (mid) at (4.25,0) {};
    	\node[goal] (1) at (5.5,0.5) {};
		\node[sink] (0) at (5.5,-0.5)  {};
		\drawdummy (bottom) at (0,-1.8) {};

		\draw[->] (init) to (p);
		\draw[->]  (p) to[bend left] node [midway,anchor=south] {$\mathsf{b}$}(q) ;
		\draw[->]  (q) to [bend left] node [midway,anchor=north] {$\mathsf{c}$} (p);
		\draw[-] (q) to node [midway,anchor=south] {$\mathsf{d}$} (mid) ;
		\draw[->] (mid) to node [below] {$\sfrac12$} (0);
		\draw[->] (mid) to node [above] {$\sfrac12$} (1);
		\draw[->]  (p) to[loop above]  node [midway,anchor=west] {$\mathsf{a}$} (0);
		\draw[->]  (0) to[loop right]  node [midway,anchor=west] {$\mathsf{f}$} (0);
		\draw[->]  (1) to [loop right] node [midway,anchor=west] {$\mathsf{e}$} (1) ;
	\end{tikzpicture}

%% file: 5_topological.tex
\section{Topological variants of sound value iteration}
\label{sec:toplogical}
The core idea of topological value iteration~\cite{TVI1} is to exploit the fact that there is an ordering among states, and certain states can never be reached again. 
More formally, a transition system can be decomposed into a directed acyclic graph of strongly connected components (SCC); the topological variant of a solution algorithm proceeds backwards through this graph, solving it component by component, see~\cite[Sec.~4.4]{MaxiGandalf-journal} for more details.
Also, we note that using the standard topological approach can in practice lead to non-termination~\cite[Sec.~4]{atva22ovi-tvi}.

We propose a new approach of exploiting the topology of the transition system, specialized for SVI.
The strength of SVI is in picking good bounds $\glb$ and $\gub$ quickly. However, the bounds are picked as the minimum/maximum over all states in $\unknown$
Thus, the global bounds are dictated by the worst-informed state in $\unknown$: as long as there exists a state with little information (e.g., high staying probability), the bounds for all states remain weak.
In particular, if there is one state with a staying probability of 1, the bounds cannot be updated at all.
These bad bounds additionally slow down convergence because the choice of strategies depends on the bounds.

To address this problem, we propose to modify the definition of the bounds, in particular the term $\opt_{s \in \unknown}~ \frac{\svireach_s^k}{1-\svistay_s^k}$ in Equation~\eqref{eq:def-bounds}, as follows:
instead of globally defining the bounds by picking the optimum among all unknown states, we define a bound for each state $s$ which only picks among those states that are reachable from $s$. Formally, let $\mathsf{All\_reachable\_states}(s)$ be the set of states reachable from $s$. Then the bounds for $s$ use the term 
\[\opt_{s' \in \unknown \cap \mathsf{All\_reachable\_states}(s)}~
\frac{\svireach_{s'}^k}{1-\svistay_{s'}^k}.\]
This is correct because the bound represents the lowest/highest possible value the state $s$ can achieve after staying for $k$ steps; and, naturally, from $s$, the play can only reach a state that is reachable from $s$.
In practice, we do not need to store a bound for every state, but instead one per SCC, because all states in the same SCC can reach each other, and thus also share their set of reachable states.

This improvement allows for more informed choices in parts of the state space where $\svireach$ and $\svistay$ already suffice to derive good bounds, without being hampered by other slow or uninformed parts of the state space. Moreover, we can stop updating an SCC when its lower and upper bound are equal, saving some resources.

%% file: 6_experiments.tex
\section{Evaluation of prototype implementation of SVI}
\label{sec:experiments}
We extended PRISM-games~\cite{prismgames3} with our prototype implementation of SVI for SG. For comparison we used the implementation of standard bounded value iteration (BVI) with deflating~\cite{KKKW18} in PRISM-games. For detailed comparison of algorithms for solving stochastic games such as value iteration (VI), optimistic VI (VI), quadric programming (QP), and strategy iteration (SI), please see our prior work~\cite{atva22ovi-tvi}.

\subsection{Technical details}
All experiments were conducted on a Linux machine with 32\,GB RAM and an Intel i7-10750H CPU at 2.60\,GHz, with a timeout of 15 minutes and a memory limit of 5\,GB. We used a precision of $\varepsilon = 10^{-6}$ throughout.

\subsection{Benchmarks}
We considered the following benchmarks:
\begin{itemize}
	\item Real-world case studies from~\cite{atva22ovi-tvi}, which subsume models from~\cite{PRISMben}.
	\item Handcrafted models from~\cite{gandalf20}, together with additional examples used to better understand the behavior of SVI.
	\item All MDP benchmarks with reachability objectives from the QVBS benchmark set~\cite{QVBS}.
\end{itemize}

\begin{figure}[t]
	\centering
	\begin{subfigure}{0.45\textwidth}
		\includegraphics[width=1\linewidth]{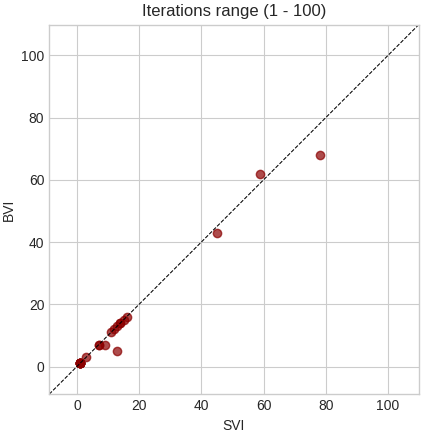}
	\end{subfigure}
	\hfill
	\begin{subfigure}{0.45\textwidth}
		\includegraphics[width=1\linewidth]{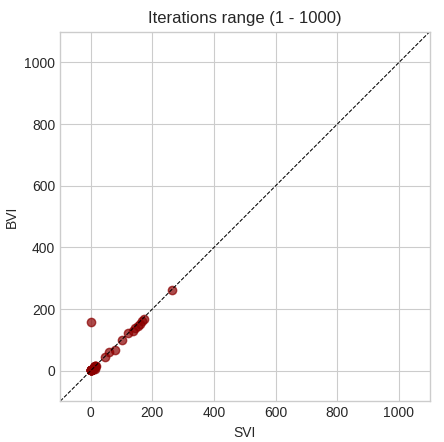}	
	\end{subfigure}

	\centering
	\begin{subfigure}{0.46\textwidth}
		\includegraphics[width=1\linewidth]{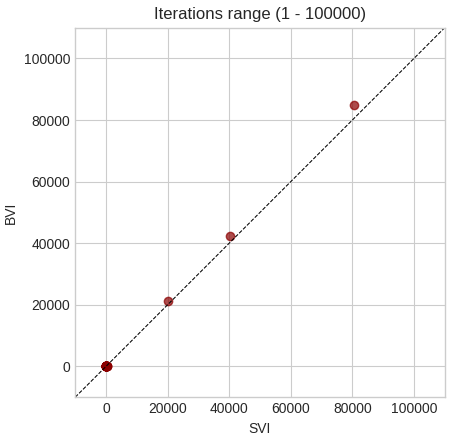}
	\end{subfigure}
	\hfill
	\begin{subfigure}{0.46\textwidth}
		\includegraphics[width=1\linewidth]{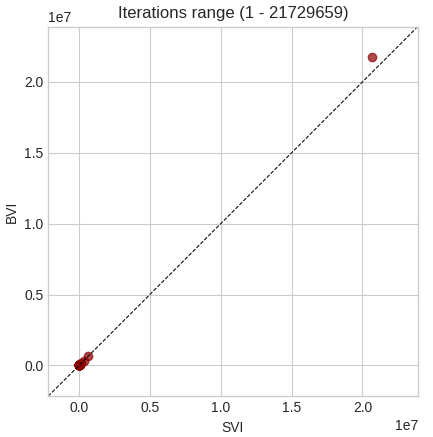}	
	\end{subfigure}
	\caption{Comparison of SVI with BVI on real and handcrafted SG benchmarks}
	\label{fig:comparisonSVI2BVI}
\end{figure}

\subsection{Algorithm comparison}

We compare the number of iterations for statistics, as the strength of SVI lies in obtaining good global bounds and resolving probabilistic loops faster. Since our implementation is a prototype, we do not focus on runtime comparisons.

Figure~\ref{fig:comparisonSVI2BVI} shows scatter plots comparing the number of iterations required by SVI and BVI on stochastic game benchmarks. Each point corresponds to one benchmark instance; points above the diagonal indicate fewer iterations for SVI, and points below indicate fewer iterations for BVI.

\begin{itemize}
	\item \textit{Comparison on SG benchmarks.}  
	We evaluated 40 instances of real and handcrafted models. SVI requires fewer iterations on 10 instances, BVI on 10 instances, and both algorithms perform equally on the remaining 20 instances.
	
	SVI performs particularly well on the \textsf{haddad-monmege}~\cite{hm18} benchmark, where it resolves probabilistic cycles efficiently. In this case, the difference in iteration counts grows with the size of the instance (e.g., 1000 $\rightarrow$ 2000 $\rightarrow$ 4000 $\rightarrow \dots$), illustrating the advantage of SVI in the presence of probabilistic loops.
	
	In contrast, BVI performs better on the \textsf{cloud} benchmark. Here, deflating allows for more effective updates across states, leading to fewer iterations for BVI. However, the difference between the two algorithms remains stable as the instance size increases.
	
	\item \textit{Comparison on MDP benchmarks.}  
	On MDP benchmarks, SVI can be applied directly in the presence of end components. We observe that SVI requires fewer iterations than BVI in most non-trivial cases (Table~\ref{tab:MDP-results}), with the exception of the \textsf{wlan\_dl} benchmark.
	
	In this benchmark, the lower decision value is set to 0 at an early stage (iteration 79), preventing further improvement of the global lower bound. As the final value is close to 1 (0.999804), this limits the effectiveness of SVI and results in a slightly higher number of iterations compared to BVI.
\end{itemize}

\label{sec:mdp_experiments}
\begin{table}
	\centering
	\caption{Comparison of SVI and BVI on MDP benchmarks where \textsf{Reach} denotes that the benchmark is solved by simple graph reachability algorithm}
	\begin{tabular}{|l|r|r|r|}
		\hline
		Model & \#States & BVI (\#Iterations) & SVI (\#Iterations) \\
		\hline
		consensus\_c1 & 528 & Reach & Reach \\
		consensus\_c2 & 1040 & 9452 & 8723 \\
		consensus\_disagree & 1040 & 10910 & 8746 \\
		csma\_all\_before\_max & 1038 & 16 & 16 \\
		csma\_all\_before\_min & 1038 & 16 & 16 \\
		csma\_some\_before & 1038 & 6 & 6 \\
		firewire\_abst & 611 & Reach & Reach \\
		firewire\_dl & 14824 & 5 & 5 \\
		firewire\_false\_elected & 4093 & Reach & Reach \\
		ij & 7 & Reach & Reach \\
		pacman & 498 & 2 & 2 \\
		philosophers & & Reach & Reach \\
		pnueli-zuck & 2701 & Reach & Reach \\
		rabin & 27766 & Reach & Reach \\
		wlan & 14883 & 275 & 271 \\
		wlan\_collisons & 2954 & Reach & Reach \\
		wlan\_dl & 542503 & 333 & 348 \\
		wlan\_sent & 2954 & Reach & Reach \\
		zeroconf\_correct\_max & 670 & 136 & 37 \\
		zeroconf\_correct\_min & 670 & 215 & 22 \\
		zeroconf\_dl\_deadline\_mn & 3835 & 44 & 28 \\
		zeroconf\_dl\_deadline\_mx & 3835 & 5 & 1 \\
		\hline
	\end{tabular}
	\label{tab:MDP-results}
\end{table}

%% file: 7_conclusion.tex
\section{Conclusion}
\label{sec:conc}
We have extended the sound value iteration from Markov decision processes to stochastic games.
In order to achieve that we had to lift the key assumption requiring that there are no end components. 
While the literature already suggests approaches to do so for other variants of value iteration, we have shown that they do not apply to sound value iteration.
Instead we had to design a new dedicated solution.
Besides, we have proposed improvements exploiting the topological properties of the models.
While the approach is mostly on par with other value iteration approaches providing guarantees on precision, there are some examples where significant speed ups can be achieved.
Yet, the point here was not a faster tool, but understanding the structure, in particular principles of evaluating cycles in SG.
The inductive, attractor-like structure of MECs provides a key theoretical foundation for handling cycles. Furthermore, it addresses a critical drawback of VI and opens the path for potential future enhancements to BVI through efficient probabilistic cycle handling.
Besides optimized implementation, the future work could include the theoretically interesting extension to concurrent stochastic games.

%% file: 8_appendix.tex
\section{Appendix: Proof of Lemma~\ref{lem:global-l-u}}
\label{sec:appendix}

\lemmagloballu*	
\begin{proof}

    \textbf{To prove (1)}, we show that the sequences $(\gub_k)_{k \ge 0}$ and $(\glb_k)_{k \ge 0}$
    are valid global upper and lower bounds, respectively.
    
    Let
    \[
    s_{\max} \in \arg\max_{s \in \unknown} \val(s)
    \qquad\text{and}\qquad
    s_{\min} \in \arg\min_{s \in \unknown} \val(s),
    \]
    and write
    \[
    \gub^* := \val(s_{\max}),
    \qquad
    \glb^* := \val(s_{\min}).
    \]
    
    We prove that for all $k \ge 0$,
    \[
    \gub_k \ge \gub^*
    \qquad\text{and}\qquad
    \glb_k \le \glb^*.
    \]
    We give the proof for the upper bound; the proof for the lower bound is dual.

	\para{Base case:}
    For $k=0$, we have $\gub_0 = 1$ by construction, and hence $\gub_0 \ge \gub^*$.
    
    \smallskip
    \para{Induction hypothesis:}
    Assume that for some $k \ge 1$, $\gub_{k-1} \ge \gub^*$.

    \para{Induction step:}
    We show that $\gub_k \ge \gub^*$.
    By Equation~\eqref{eq:def-bounds},
    \[
    \gub_k
    =
    \min\!\left(
    \gub_{k-1},
    \max\!\left(
    \max_{s \in \unknown} \frac{\svireach^k_s}{1-\svistay^k_s},
    \ \svidecMax
    \right)
    \right).
    \]
    We distinguish cases according to this definition.

    \noindent\emph{Case 1:}
    $\gub_k = \gub_{k-1}$.

    Then, by the induction hypothesis, $\gub_{k-1} \ge \gub^*$, and hence $\gub_k \ge \gub^*$.

	\medspace
	
    \para{Auxiliary construction:} We first establish a key inequality for the state $s_{\max}$ that we use to prove \emph{Case 2} and \emph{Case 3}.

    For the strategies $\sigma_k$ and $\tau_k$, we have
    \begin{align}
    	\probability_{s_{\max}}^{\sigma_k, \tau_k} (\reach^{\leq k} \targets)
    	+
    	\probability_{s_{\max}}^{\sigma_k, \tau_k} (\stay^{\leq k} \unknown) \cdot \gub_{k-1}
    	\geq \gub^*
    	\tag{by Theorem~\ref{thm:svi-mc}}
    \end{align}

    Using Lemma~\ref{lem:svi:inv-1}, this can be rewritten as
    \begin{align}
    		\label{eq:upperbound-s-max}
    		\svireach_{s_{\max}}^k + \svistay_{s_{\max}}^k \cdot \gub_{k-1} \geq \gub^*
    \end{align}

    Define the sequence $(v_i)_{i \ge 0}$ by
    \begin{align*}
    	v_0 &= \gub_{k-1},\\
    	v_{i+1} &= f(v_i)
    	= \svireach^k_{s_{\max}} + \svistay^k_{s_{\max}} \cdot v_i.
    \end{align*}
    
    Since $\svireach^k_{s_{\max}}$ and $\svistay^k_{s_{\max}}$ are fixed,
    $f$ is an affine function on $[0,1]$.
    As $\svistay^k_{s_{\max}} < 1$ (from Algorithm~\ref{alg:svi-sg-high-level}), the function $f$ has a unique fixed point.
    Hence,
	\begin{align}
		\label{eq:v_infty}
		v_\infty \eqdef \lim_{i \to \infty} v_i = \dfrac {\svireach^k_{s_{\max}}} {1 - \svistay^k_{s_{\max}}}
	\end{align}

	\smallskip
	\para{} Moreover, we have the following:
    \begin{claim}
	\label{eqn:v_i}
	For all $i \geq 0$, we have $v_{i+1} \leq v_i$.
    \end{claim}
    
    \begin{proof}
    Since $s_{\max} \in \unknown$, Algorithm~\ref{alg:svi-sg-high-level} ensures that
    \[
    \svistay^k_{s_{\max}} < 1.
    \]
    Moreover, by definition of $f$ and of the fixed point $v_\infty$, we have
    \[
    f(v_\infty)=v_\infty.
    \]
    Now,
    \begin{align}
    	v_{i+1}
    	&= f(v_i) \nonumber\\
    	&= f(v_\infty + v_i - v_\infty)
    	\tag{adding and subtracting $v_\infty$}\nonumber\\
    	&= \svireach^k_{s_{\max}} + \svistay^k_{s_{\max}} \cdot (v_\infty + v_i - v_\infty)
    	\tag{by definition of $f$}\nonumber\\
    	&= \svireach^k_{s_{\max}} + \svistay^k_{s_{\max}} \cdot v_\infty
    	+ \svistay^k_{s_{\max}} \cdot (v_i - v_\infty)\nonumber\\
    	&= f(v_\infty) + \svistay^k_{s_{\max}} \cdot (v_i - v_\infty)
    	\tag{by definition of $f$}\nonumber\\
    	&= v_\infty + \svistay^k_{s_{\max}} \cdot (v_i - v_\infty)
    	\tag{since $v_\infty$ is a fixed point of $f$}\nonumber\\
    	&\leq v_\infty + (v_i - v_\infty)
    	\tag{since $\svistay^k_{s_{\max}} < 1$}\nonumber\\
    	&= v_i.\nonumber
    \end{align}
    This proves the claim.
    \end{proof}

    \begin{claim}
	\label{eqn:v_i-update-donot-overshoot}
	If $v_i \in [\svidecMax,\gub_{k-1}]$, then $v_{i+1} \ge \gub^*$.
    \end{claim}
    
    \begin{proof}
    By Equation~\eqref{eq:upperbound-s-max}, we have
    \[
    \svireach_{s_{\max}}^k + \svistay_{s_{\max}}^k \cdot \gub_{k-1} \ge \gub^*.
    \]
    
    Moreover, by definition of the decision value $\svidecMax$, replacing $\gub_{k-1}$ by any
    value in the interval $[\svidecMax,\gub_{k-1}]$ preserves this inequality. Since
    $v_i \in [\svidecMax,\gub_{k-1}]$, it follows that
    \[
    \svireach_{s_{\max}}^k + \svistay_{s_{\max}}^k \cdot v_i \ge \gub^*.
    \]
    
    By definition of $v_{i+1}$, this is exactly
    \[
    v_{i+1} \ge \gub^*.
    \]
    \end{proof}
    	
    \medspace
    
    \noindent\emph{Case 2:}
    $\gub_k = \svidecMax$.
    
    We show that $\svidecMax \ge \gub^*$.
    
    Since
    \[
    \gub_k
    =
    \min\!\left(
    \gub_{k-1},
    \max\!\left(
    \max_{s \in \unknown} \frac{\svireach^k_s}{1-\svistay^k_s},
    \ \svidecMax
    \right)
    \right)
    \]
    
    and $\gub_k = \svidecMax$, it follows that
    \[
    \svidecMax \ge \max_{s \in \unknown} \frac{\svireach^k_s}{1-\svistay^k_s}.
    \]
    
    In particular,
    \[
    \svidecMax \ge \frac{\svireach^k_{s_{\max}}}{1-\svistay^k_{s_{\max}}}
    = v_\infty.
    \]
    
    Moreover, $v_0 = \gub_{k-1}$, and by Claim~\ref{eqn:v_i} the sequence $(v_i)_{i\geq 0}$
    is monotonically non-increasing and converges to $v_\infty$. Hence all iterates satisfy
    \[
    v_i \in [v_\infty,\gub_{k-1}].
    \]
    
    Since $\svidecMax \ge v_\infty$, there exists some index $i$ such that
    \[
    v_i \in [\svidecMax,\gub_{k-1}].
    \]
    
    By Claim~\ref{eqn:v_i-update-donot-overshoot}, this implies
    \[
    v_{i+1} \ge \gub^*.
    \]
    
    As $(v_i)_{i\geq 0}$ converges to $v_\infty$, we obtain
    \[
    v_\infty \ge \gub^*.
    \]
    
    Together with $\svidecMax \ge v_\infty$, this yields
    \[
    \svidecMax \ge \gub^*.
    \]
	
	\medspace

    \noindent\emph{Case 3:}
    \(
    \gub_k=\max_{s\in\unknown} \frac{\svireach^k_s}{1-\svistay^k_s}.
    \)
    
    We show that
    \[
    \max_{s\in\unknown} \frac{\svireach^k_s}{1-\svistay^k_s} \ge \gub^*.
    \]
    
    Since $s_{\max}\in\unknown$, it is enough to prove
    \[
    \frac{\svireach^k_{s_{\max}}}{1-\svistay^k_{s_{\max}}} \ge \gub^*.
    \]
    
    By the definition of $v_\infty$ in Equation~\eqref{eq:v_infty}, this is equivalent to showing
    \[
    v_\infty \ge \gub^*.
    \]
    
    From Equation~\eqref{eq:def-bounds}, we have
    \[
    \gub_{k-1}
    \ge
    \max_{s\in\unknown} \frac{\svireach^k_s}{1-\svistay^k_s}
    \ge
    \frac{\svireach^k_{s_{\max}}}{1-\svistay^k_{s_{\max}}}
    =
    v_\infty.
    \]
    
    Hence $v_0=\gub_{k-1}\ge v_\infty$.
    
    Moreover, since $f(v)=\svireach^k_{s_{\max}}+\svistay^k_{s_{\max}}\cdot v$ is monotone,
    from $v_i\ge v_\infty$ we obtain
    \[
    v_{i+1}=f(v_i)\ge f(v_\infty)=v_\infty.
    \]
    
    Thus, by induction, $v_i\ge v_\infty$ for all $i\ge 0$.
    
    By Claim~\ref{eqn:v_i}, the sequence $(v_i)_{i\ge 0}$ is monotonically non-increasing.
    Since it converges to $v_\infty$, every term satisfies
    \[
    v_i\in [v_\infty,\gub_{k-1}].
    \]
    
    Because in the present case
    \[
    \max_{s\in\unknown} \frac{\svireach^k_s}{1-\svistay^k_s} \ge \svidecMax,
    \]
    
    we have in particular $v_\infty\ge \svidecMax$. Therefore every $v_i$ lies in
    $[\svidecMax,\gub_{k-1}]$, and Claim~\ref{eqn:v_i-update-donot-overshoot} yields
    \[
    v_{i+1}\ge \gub^*
    \qquad\text{for all } i\ge 0.
    \]
    
    Passing to the limit and using $v_i\to v_\infty$, we obtain
    \[
    v_\infty\ge \gub^*.
    \]
    
    Therefore,
    \[
    \gub_k
    =
    \max_{s\in\unknown} \frac{\svireach^k_s}{1-\svistay^k_s}
    \ge
    \frac{\svireach^k_{s_{\max}}}{1-\svistay^k_{s_{\max}}}
    =
    v_\infty
    \ge
    \gub^*.
    \]
    This completes the proof of (1).

	\medspace
    
    \noindent \para{To prove (2)},
    we again consider the three cases arising from the definition of $\gub_k$ and show that
    \(
    \gub_k \ge \max_{s\in\unknown} \probability_s^{\sigma_k,\tau_k}(\reach \targets).
    \)
    The proof for the lower bound is dual.

	\medspace

    \noindent \emph{Case 1:} $\gub_k=\gub_{k-1}$.
    
    We show that
    \[
    \gub_{k-1} \ge \max_{s\in\unknown} \probability_s^{\sigma_k,\tau_k}(\reach \targets).
    \]
    
Taking the maximum over all $s \in \unknown$ yields the claim.
	
	For all $s \in \unknown$, we have:
	\begin{align}
		\gub_{k-1} &\geq \probability_{s}^{\sigma_{k}, \tau_k} (\reach^{\leq k} \targets) + \probability_{s}^{\sigma_{k}, \tau_k} (\stay^{\leq k} \unknown) \cdot \gub_{k-1} \tag{By Argument 1} \nonumber\\
		&\geq \probability_{s}^{\sigma_{k},\tau_{k}}(\reach \targets) \tag{By Theorem~\ref{thm:svi-mc}}\nonumber\\
		&\geq \max_{s\in\unknown} \probability_{s}^{\sigma_{k},\tau_{k}}(\reach \targets)\tag{By pushing universal quantification}
	\end{align}
	
	\para{Argument 1:} First inequality holds because $\gub_{k-1}$ is correct upperbound on the value of all states of the underlying Markov chain under the strategies $\sigma_{k}$ and $\tau_{k}$.

    \medspace

    \noindent \emph{Case 2:}
    \(\gub_k=\max_{s\in\unknown} \frac{\svireach^k_s}{1 - \svistay^k_s}.\)

    We show that
    \[
    \max_{s\in\unknown} \frac{\svireach^k_s}{1 - \svistay^k_s}
    \ge
    \max_{s\in\unknown} \probability_s^{\sigma_k,\tau_k}(\reach \targets).
    \]
    
    Let $s_k^{\max} \in \arg\max_{s\in\unknown}
    \probability_s^{\sigma_k,\tau_k}(\reach \targets)$.
    It suffices to prove
    \[
    \frac{\svireach^k_{s_k^{\max}}}{1 - \svistay^k_{s_k^{\max}}}
    \ge
    \probability_{s_k^{\max}}^{\sigma_k,\tau_k}(\reach \targets).
    \]
    
    By Theorem~\ref{thm:svi-mc}, we have
    \[
    \probability_{s_k^{\max}}^{\sigma_k,\tau_k}(\reach \targets)
    \le
    \probability_{s_k^{\max}}^{\sigma_k,\tau_k}(\reach^{\leq k} \targets)
    +
    \probability_{s_k^{\max}}^{\sigma_k,\tau_k}(\stay^{\leq k} \unknown)
    \cdot
    \probability_{s_k^{\max}}^{\sigma_k,\tau_k}(\reach \targets).
    \]
    
    Rearranging yields
    \[
    \probability_{s_k^{\max}}^{\sigma_k,\tau_k}(\reach \targets)
    \le
    \frac{
    \probability_{s_k^{\max}}^{\sigma_k,\tau_k}(\reach^{\leq k} \targets)
    }{
    1 -
    \probability_{s_k^{\max}}^{\sigma_k,\tau_k}(\stay^{\leq k} \unknown)
    }.
    \]
    
    Taking the maximum over all $s \in \unknown$ gives the claim.

    \medspace
    
	\noindent \emph{Case 3:}
    \(\gub_k=\svidecMax.\)
    
    We show that
    \[
    \svidecMax \ge \max_{s\in\unknown} \probability_s^{\sigma_k,\tau_k}(\reach \targets).
    \]
    
    By the definition of $\gub_k$, we have
    \[
    \svidecMax \ge \max_{s\in\unknown} \frac{\svireach^k_s}{1-\svistay^k_s}.
    \]
    
    By the argument from Case 2,
    \[
    \max_{s\in\unknown} \frac{\svireach^k_s}{1-\svistay^k_s}
    \ge
    \max_{s\in\unknown} \probability_s^{\sigma_k,\tau_k}(\reach \targets).
    \]
    
    Combining the two inequalities yields
    \[
    \svidecMax \ge \max_{s\in\unknown} \probability_s^{\sigma_k,\tau_k}(\reach \targets).
    \]
	
	\para{Summary of the proof.}
    We have shown that for all $k \ge 0$, the value $\gub_k$ is a valid global upper bound.
    The decision value $\svidecMax$ is essential in Cases 2 and 3: it prevents the update from
    dropping below the threshold induced by the fixed-point value
    \(
    \frac{\svireach^k_s}{1 - \svistay^k_s},
    \)
    thereby ensuring that the invariant $\gub_k \ge \gub^*$ is preserved.

    The proof for $\glb_k$ is analogous and shows that $\glb_k$ is a valid global lower bound.
	
\end{proof}